\documentclass[11pt]{article}
\pdfoutput=1
\usepackage{jheppub}
\usepackage{graphicx}
\usepackage{amssymb}
\usepackage{amsmath,amssymb}
\usepackage{slashed}
\usepackage{hyperref}
\usepackage{caption}
\usepackage{xcolor}
\usepackage{dsfont}
\usepackage{verbatim}
\usepackage{subfig}
\usepackage{pgfplots,amsfonts, xspace} 
\usetikzlibrary{arrows,shapes,positioning,calc,decorations.pathmorphing,decorations.markings}
\DeclareMathSymbol{\mlq}{\mathrel}{operators}{``}
\DeclareMathSymbol{\mrq}{\mathrel}{operators}{`'}
\DeclareMathSymbol{\mlqq}{\mathrel}{operators}{"5C}
\DeclareMathSymbol{\mrqq}{\mathrel}{operators}{`"}

\newcommand\beq{\begin{equation}}
\newcommand\eeq{\end{equation}}
\newcommand\be{\begin{equation}}
\newcommand\ee{\end{equation}}

\title{Critical theories connecting gapped phases with $\mathbb{Z}_2\times\mathbb{Z}_2$ symmetry from the duality web}

\preprint{\today}

\author[a]{Andreas Karch, Ryan C. Spieler}
\affiliation[a]{University of Texas, Austin, Physics Department, Austin, TX, 78712, USA}
\emailAdd{karcha@utexas.edu,rcspieler@utexas.edu}

\abstract{We use the ideas  behind the duality web to construct numerous conformal field theories mediating the phase transitions between various symmetry broken and topological phases. In particular we obtain the full field theory version of the Kennedy Tasaki transformation, mapping a gapless theory mediating a topological phase transition of symmetry protected topological orders to a standard symmetry breaking one in a 1+1 dimensional $\mathbb{Z}_2 \times \mathbb{Z}_2$ gauge theory. When we consider all possible discrete gauging operations, we obtain bosonic and fermionic webs with 9 critical theories  per web, each connecting 4 separate gapped phases, some of them topological. Bosonization maps the two webs into each other.  In addition to discussing the multi-critical theory connecting the four gapped phases in each phase diagram, we discuss the partially gapped theories connecting two of those four.  Some of these are gapless symmetry protected topological phases.}

\begin{document}
\maketitle

\section{Introduction}

The idea of the duality web \cite{Karch:2016sxi,Seiberg:2016gmd} was first introduced in 2+1 dimensional Chern-Simons matter theories with $U(1)$ gauge symmetries, using formal path integral identities as had previously been employed for supersymmetric dualities \cite{Kapustin:1999ha}. A simple seed duality, when read as an exact equivalence of the partition function of two quantum field theories as a function of background fields, can be used to generate a large web of dualities by repeated promotion of backgrounds to dynamical fields. In particular, using 3d bosonization \cite{Polyakov:1988md,Shaji:1990is,Paul:1990vw,Fradkin:1994tt,Chen:1993cd,Barkeshli:2012rja} as a seed, one can derive a large class of 2+1 dimensional dualities, including the standard particle vortex duality \cite{Peskin:1977kp,Dasgupta:1981zz} and its fermionic cousin \cite{Son:2015xqa}.

A similar web of dualities can be found in 1+1 dimensions \cite{Karch:2019lnn} as hinted at in \cite{Senthil:2018cru}. Here the standard Jordan-Wigner (JW) bosonization/fermionization transformation \cite{Jordan:1928wi} serves as the seed from which many other famous dualities can be derived, including the Kramers-Wannier \cite{Kramers:1941kn} duality of the Ising model as well as T-duality of string theory. While many of these dualities had first been formulated in lattice systems, one can view them as precise equivalences of continuum quantum field theories. In particular the JW seed duality can either be read as an equivalence of a Majorana fermion coupled to a $\mathbb{Z}_2$ gauge field and a self-interacting boson, or that of a free Majorana fermion and a boson coupled to a $\mathbb{Z}_2$ gauge field. Once one has one of the two versions established, the other one follows \cite{Karch:2019lnn}. By repeated application of the seed one can derive a large web of dualities involving $\mathbb{Z}_2$ symmetries, both gauged and global.

Another famous example of a 1+1 dimensional transformation involving $\mathbb{Z}_2$ symmetries was introduced by Kennedy and Tasaki \cite{Kennedy:1992ifl,Kennedy:1992tke} (KT). In the language of the web, this is not a duality in the sense of an exact equivalence of partition functions, but rather it is a well defined gauging procedure that takes one theory into a different one in a controlled way, so that the partition function of the latter can be obtained from the former, but is not equal to it. These kind of transformations generate the web. What has brought the KT transformation into renewed focus recently is that it maps the phase transition between a trivial and non-trivial $\mathbb{Z}_2 \times \mathbb{Z}_2$ symmetry protected topological (SPT) phase to a standard spontaneous symmetry breaking transition. This is an important ingredient in recent attempts to map all phase transitions to standard Landau symmetry breaking ones. While the KT transformation is non-local, it maps local symmetry algebras into local symmetry algebras. Our goal is to demonstrate that KT fits neatly into the duality web. While the original KT transformation was constructed in a lattice model, we will use its continuum version as laid out in \cite{Li:2023ani}. Furthermore, while \cite{Li:2023ani} only considered the almost trivial field theory describing the gapped phases, we apply the KT transformation to the conformal field theory mediating the phase transition from which the gapped phases can be obtained by relevant deformations. The conformal field theory describing the standard symmetry breaking phase transition is of course well known. Acting with the KT transformation on this CFT gives a CFT mediating the transition between the SPT phase and the trivial phase which turns out to be just a free boson\footnote{This CFT has previously been studied in \cite{Tsui:2017ryj}, but we believe the explicit Lagrangian realization we get using KT makes its properties much more explicit.}\footnote{This calculation was also done in \cite{Cordova:2022lms}, which the authors found out about after posting this work.  We also want to point out field theoretic studies of the KT transformation in i.e. \cite{Li:2023knf,Bhardwaj:2023bbf,Bhardwaj:2024ydc}.}. We show that this structure is part of a web connecting 9 different CFTs, each with 2 relevant deformations and hence generating a phase diagram with 4 different phases. The CFT sits at the origin. As we discuss in detail, the axes generically correspond to topologically gapless phases.

We perform a similar analysis for fermionic theories with $\mathbb{Z}_2^F \times \mathbb{Z}_2$ global symmetry and, once again,  construct 9 different CFTs, each one sitting at the origin of a 2d phase diagram realizing 4 separate gapped phases. The bosonic and fermionic webs of CFTs can be mapped into each other via fermionization/bosonization.

We'll discuss the bosonic web in section 2, and the fermionic web, as well as the map between the two, in section 3. We present the complete set of Lagrangians and a detailed analysis of the phase structures in the appendices.

\section{The Web of Bosonic CFTs with $\mathbb{Z}_2 \times \mathbb{Z}_2$ symmetry}
In this section, we discuss the web of bosonic CFTs generated by various discrete gauging operations in theories with a non-anomalous $\mathbb{Z}_2\times\mathbb{Z}_2$ symmetry.  We begin by reviewing the field-theoretic description of the Kennedy-Tasaki transformation between SSB and SPT phases.  We then apply this transformation directly to $\text{Ising}^2$ CFT, showing that it governs the desired phase transition between the trivial and SPT phases.  We then study all discrete gauging operations and the phase diagrams and CFTs they generate before discussing the partially gapped theories that appear in those phase diagrams.
\subsection{Kennedy-Tasaki in the Continuum}

The original Kennedy-Tasaki transformation is an operation on gapped systems with $\mathbb{Z}_2 \times \mathbb{Z}_2$ global symmetry. As such, the relevant partition functions depend only on background fields, which in the case at hand are two $\mathbb{Z}_2$ background gauge fields $A_1$ and $A_2$. According to \cite{Li:2023ani} the three options for the partition function that are acted on are the topological trival phase
\beq Z_{\text{Tri}}[A_1,A_2] = 1, \eeq
the spontaneous symmetry breaking phase
\beq Z_{\text{SSB}}[A_1,A_2] = \delta(A_1) \delta(A_2), \eeq
and the non-trivial SPT phase 
\beq Z_{\text{SPT}}[A_1,A_2] = (-1)^{\int A_1  A_2} \eeq
where the exponent is short hand for the standard topological cup product for $\mathbb{Z}_2$ gauge fields akin to a CS term for gauge fields in 2+1 dimensions.
According to \cite{Li:2023ani} we can act on these partition functions $Z_I$, where $I$ stands for Tri, SSB, and SPT, with two topological manipulations termed\footnote{In \cite{Li:2023ani} the sum over gauge fields in the $S$ transformation contains an extra prefactor of 
$1/|H^0(X_2,\mathbb{Z}_2)|$. As explained in \cite{Kaidi:2022cpf}, the normalization of these sums can be altered by a power of
$$ \chi[x_2,\mathbb{Z}_2]
= \frac{|H^0(X_2,\mathbb{Z}_2)||H^2(X_2,\mathbb{Z}_2)|}{|H^1(X_2,\mathbb{Z}_2)|}.
$$
by adding an extra Euler counterterm to the action. A very convenient choice is to chose the $(-1/2)$-th power of the Euler character, which yields a prefactor of $1/\sqrt{|H^1(X_2,\mathbb{Z}_2)|}$, which for a genus $g$ Riemann surface is simply $2^{-g}$. This is also the normalization used in \cite{Karch:2019lnn}. Furthermore, for simplicity we chose not to explicitly display these factors; all our sums over gauge fields are implicitly normalized with this prefactor, and so are the delta functions. That is, when we write $\sum_a$ we mean a sum over the flat connections with a prefactor of $2^{-g}$, and $$\delta(A) = 2^g \delta_{A0},$$ where the second delta is the Kronecker Delta for the discrete choices of connection. This conventions ensures that $$\sum_{ab} (-1)^{\int ab} =1.$$ 
. Note that the factor of $2^g$ in the delta function is indeed needed to correctly account for the 2-fold degeneracy of the groundstate in the symmetry broken phase.} $S$ and $T$:
\begin{eqnarray}
Z_I[A_1,A_2] &\xrightarrow[]{\, \,  S \, \,}&  \sum_{a_1,a_2} Z_I[a_1,a_2] \, (-1)^{ \int a_1  A_2 + a_2 A_1} \\
Z_I[A_1,A_2] &\xrightarrow[]{\, \,  T \, \,}&  Z_I[A_1,A_2] \, (-1)^{ \int A_1  A_2 }.
\end{eqnarray}
$S$ amounts to a non-trivial gauging of the two $\mathbb{Z}_2$ symmetries, the second stacks a non-trivial SPT. These are once again the straightforward generalizations of the corresponding $S$ and $T$ transformations employed in 2+1 dimensions CS gauge theories \cite{Witten:2003ya}. They act on the partition functions as follows:
\begin{eqnarray}
T Z_{\text{SSB}} = Z_{\text{SSB}}, &\quad& S Z_{SSB} = Z_{\text{Tri}} \\
T Z_{\text{Tri}} = Z_{\text{SPT}}, &\quad& S Z_{Tri} = Z_{\text{SSB}} \\
T Z_{\text{SPT}} = Z_{\text{Tri}}, &\quad& S Z_{SPT} = Z_{\text{SPT}} .
\end{eqnarray}
In terms of these the KT transformation is
\beq
KT = STS = TST.
\eeq
In particular, $STS=TST$ leaves the trivial phase untouched, but turns SPT into SSB and vice versa.

\subsection{Kenneday-Tasaki for Gapless Theories}

What we are interested in here is not just KT as a transformation between isolated gapped phases, but as an action on the phase diagram including the conformal field theory (CFT) mediating the second order phase transition between them. $Z_{\text{tri}}$ and $Z_{\text{SSB}}$ are connected by a completely standard Landau symmetry breaking phase transition with a CFT, which we call CFT$_{SSB}$, capturing the critical fluctuations of the order parameter. KT maps the gapped phases on either side of the transition to trival and SPT phase respectively and so should map CFT$_{SSB}$ to a new CFT, CFT$_{SPT}$, which mediates the phase transition between these two phases. Our goal is to use the duality web techniques to write down these two CFTs and demonstrate that they are indeed mapped into each other using $STS$ \footnote{After submitting this paper, the authors became aware of \cite{Cordova:2022lms}, which does this calculation in its section 5.}.

In equations, we are looking for two CFTs with the following properties
\beq
Z_{CFT_{SSB}} \xrightarrow[]{m} \left \{ \begin{array}{lll} Z_{\text{SSB}}& \mbox{ for }& m>0 \cr 
Z_{\text{Tri}} &\mbox{ for }& m<0  \end{array} \right . , \quad \quad \quad
 Z_{CFT_{SPT}} \xrightarrow[]{m} \left \{ \begin{array}{lll} Z_{\text{SPT}}& \mbox{ for }& m>0 \cr 
Z_{\text{Tri}} &\mbox{ for }& m<0  \end{array} \right . \label{property}
\eeq
with
\beq
TST \,  Z_{CFT_{SSB}} = Z_{CFT_{SPT}}.
\eeq
Here $m$ denotes a relevant deformation that drives the respective CFT into either one of the allowed gapped phases, depending on sign.

In \cite{Lu:2024ytl} it was pointed out that there is a second combination of $S$ and $T$ that can be used to map gapped phases and CFTS into each other: the triality operator $ST$. Unlike KT, which is an element of order two (meaning it squares to the identity), the triality operator is of order three and permutes the 3 gapped phases. The triality operation also gives rise to a third CFT, the one mediating the transition between SSB and SPT. We will return to this in the next subsection when we discuss the full bosonic duality web.

Clearly the simplest version for CFT$_{SSB}$ is just two copies of the standard Ising CFT. This theory is equivalent to a single boson with target space $S^1/\mathbb{Z}_2$, which has a marginal deformation: the size of the interval. In this work we will restrict ourselves to the point where the theory fermionizes into two free Majorana fermions coupled to a $\mathbb{Z}_2$ gauge field. The marginal deformation comes along for the ride under all the transformations we apply; every single CFT we write is an entire line worth of CFTs. This marginal deformation plays an important role in the work of \cite{Lu:2024ytl}, where it was argued that the marginal deformation of the three CFTs connected by triality in the end lead into a single CFT that is triality invariant, the Kosterlitz-Thouless point. 

Returning to the two copies of the Ising model, we write the partition function as
\beq Z_{CFT_{SSB}}[A_1,A_2] = Z_{Is}[A_1] + Z_{Is}[A_2] \eeq
where
\beq Z_{Is}[A] = \int {\cal D \phi}\,  e^{i S_{Is}[\phi,A]} \eeq
and the action is written in the form used in \cite{Karch:2019lnn}
\beq
S_{Is}[\phi,A] = \int  (D_A \phi)^2 + \phi^4 .
\eeq
Here $D_A$ is a covariant derivative coupling the dynamical Ising scalar $\phi$ to the background $\mathbb{Z}_2$ gauge field $A$. The relevant deformation we can add for either of the two Ising scalars with parameter $M^2$ is the mass squared of the scalar $\phi$. The phase transition we are looking at is multi-critical in that first we dial the mass squareds of both scalars simultaneously. We will look at the full phase diagram with unequal masses in the next subsection.

Our task is to demonstrate that acting with $TST$ on this CFT$_{SSB}$ indeed yields a CFT$_{SPT}$ with the property \eqref{property}. Having laid out our conventions we are in a position to explicitly apply $TST$ on CFT$_{SSB}$. To do so we explicitly work out the impact of the various transformations on the field theory. The original action is given by ($j=1,2$)
\beq S_{CFT_{SSB}} = \int \sum_j [ (D_{A_j} \phi_j)^2 + \phi_j^4 ] .\eeq
Acting with $T$ we get
\beq S_{T } =  \int\sum_j [ (D_{A_j} \phi_j)^2 + \phi_j^4]  + i \pi A_1 A_2 .\eeq
Acting with $S$ on this we introduce two more dynamical fields $a_1$ and $a_2$ and get the action
\beq S_{ST}= \int \sum_j [ (D_{a_j} \phi_j)^2 + \phi_j^4]  + i \pi (a_1 a_2 + a_1 A_2 + a_2 A_1) .\eeq
Last but not least acting with one more $T$ we finally arrive at
\beq S_{TST}= \int \sum_j [ (D_{a_j} \phi_j)^2 + \phi_j^4]  + i \pi (a_1 a_2 + a_1 A_2 + a_2 A_1 + A_1 A_2) .\eeq
By construction, this CFT we obtain by acting on the two copies of the Ising CFT mediating the standard symmetry breaking phase transition should be the correct CFT describing the transition between trivial and SPT phase, that his $S_{TST}$ {\it is} supposed to be the action of CFT$_{SPT}$. To confirm that his is actually the case,  we should show that the CFT described by the action $S_{TST}$ has the desired properties of CFT$_{SPT}$ from \eqref{property} above.

Let us study the phases one at a time:

\noindent \underline{$M^2>0$}: 
If we give the scalars a positive mass squared they simply decouple from the theory. We are left with a purely topological action:
\beq S_{M^2>0}= i \pi \int (a_1 a_2 + a_1 A_2 + a_2 A_1 + A_1 A_2) .\eeq
We can integrate out, say, $a_1$, setting $a_2 = A_2$. That leaves us behind with two copies of the cup product, which is indeed equivalent to the trivial theory.

\noindent \underline{$M^2<0$}: 
If we give the scalars a negative mass squared we Higgs the $\mathbb{Z}_2$ gauge groups, that is the path integral over the scalars produces factors of $\delta(a_1) \delta(a_2)$ in the partition function. This sets 3 of the four cup products in the action to zero, leaving behind a single $A_1 A_2$ term, and hence the resulting theory is indeed the non-trivial SPT phase.

Having verified that CFT$_{SPT}$ has the right phase structure, let us end by noting that this CFT can in fact be rewritten in a much simpler form: it is just a free boson at radius 2! This is not obvious in the form we write the action, but it has in fact been shown in \cite{Karch:2019lnn} that this is indeed the case. In there this same action was obtained as a $\mathbb{Z}_2$ gauging of a Dirac fermion. As this is a CFT with central charge $c=1$ it had to end up being equivalent to one of the known bosonic theories. The authors of \cite{Karch:2019lnn} use duality arguments to demonstrate that it corresponds to a compact boson at radius 2, and verified some of its expected properties (existence of a marginal deformation that corresponds to changing the radius, T-duality, and an enhanced global symmetry at $R=\sqrt{2}$). In this description of the theory as a free boson, it is also obvious how to calculate all correlation functions.

\subsection{The Full Web of Bosonic Phase Diagrams}

So far we have focused on the case where the mass squareds of the two Ising scalars (or the mass of the two Majorana fermions after applying KT) are always dialed together. The full phase diagrams of these theories is two dimensional, parametrized by independently dialing the coefficients of the two relevant deformations\footnote{Some discussion of this appeared in \cite{Lichtman:2020nuw,Chatterjee:2022tyg,Moradi:2022lqp,Huang:2023pyk}, but they do not seem as direct or exhaustive as ours.}. For completeness, let us work out the full 2d phase diagrams that can be realized this way and how they are connected by the various gauging operations. There are five nontrivial gapped phases with $\mathbb{Z}_2\times\mathbb{Z}_2$ global symmetry.  The first, which we continue to call SSB, spontaneously breaks the entire $\mathbb{Z}_2\times\mathbb{Z}_2$ symmetry.  The second two, SSB1 and SSB2, break the $\mathbb{Z}_2\times\mathbb{Z}_2$ symmetry to one of its factors.  The fourth, SSBD, breaks the $\mathbb{Z}_2\times\mathbb{Z}_2$ symmetry to its diagonal subgroup.  The fifth, SPT, is the nontrivial SPT phases that we used above to implement the Kennedy-Tasaki transformation.  The partition functions for these phases are:
\begin{eqnarray}
 Z_{SSB} &=& \delta(A_1)\delta(A_2)\\
 Z_{SSB1} &=& \delta(A_1) \\
 Z_{SSB2} &=& \delta(A_2) \\
 Z_{SSBD} &=& \delta(A_1+A_2) \\
 Z_{SPT} &=& (-1)^{\int A_1A_2}
\end{eqnarray}
There are three operations that we can use to map among the gapped phases.  We can gauge the first copy of $\mathbb{Z}_2$, gauge the second copy of $\mathbb{Z}_2$, or stack with the nontrivial SPT phase.  Following \cite{Gaiotto:2020iye}, we refer to these operations as $O_1$, $O_2$, and $S_1$, respectively.  Their actions on a partition function $Z[A_1,A_2]$ are
\beq
Z[A_1,A_2] \xrightarrow[]{\, \,  O_1 \, \,} \sum_{a_1} Z[a_1,A_2] \, (-1)^{\int a_1 A_1},
\eeq
\beq
Z[A_1,A_2] \xrightarrow[]{\, \,  O_2 \, \,} \sum_{a_2} Z[A_1,a_2] \, (-1)^{\int a_2 A_2},
\eeq
and
\beq
Z[A_1,A_2] \xrightarrow[]{\, \,  S_1 \, \,} Z[A_1,A_2] \, (-1)^{\int A_1 A_2}.
\eeq
Note that these include the $S$ and $T$ operations above, since $S = O_1 O_2$ \footnote{Note that we swapped the labels 1 and 2 on the background fields for the dual symmetry compared to the above.} and $T=S_1$.

The action of $O_1$, $O_2$, and $S_1$ on these theories is simple to work out. One delta function is sufficient to set the putative $\int A_1 A_2$ topological term to 0, so $S_1$ leaves these partial SSB phases invariant, whereas the gauging in $O_1$ and $O_2$ removes the existing delta function but introduces the other one, permuting SSB, SSB1, and SSB2.  The exception is SSBD, which is mapped to SPT by $O_1$ and $O_2$. All in all the transformations act as displayed in table \ref{bosonicphases}\footnote{To derive some of these and results in Appendices \ref{aA1} and \ref{aA2}, the reader should recall that the cup product is supercommutative on cohomology so that terms like $a^2$ vanish.  This assumes our spacetime to be orientable.}.

\begin{table}
\begin{center}
\begin{tabular}{|c|c|c|c|c|c|c|}
\hline
Original & Tri & SSB&SSB1&SSB2&SSBD&SPT \\
\hline
$O_1$ on Original &SSB1 & SSB2&Tri&SSB&SPT&SSBD \\
\hline
$O_2$ on Original &SSB2&SSB1&SSB&Tri&SPT&SSBD \\
\hline
$S_1$ on Original & SPT & SSB & SSB1 &  SSB2 & SSBD & Tri\\
\hline
\end{tabular}
\caption{Action of $O_1$, $O_2$ and $S_1$ on the gapped bosonic phases.}
\label{bosonicphases}
\end{center}
\end{table}

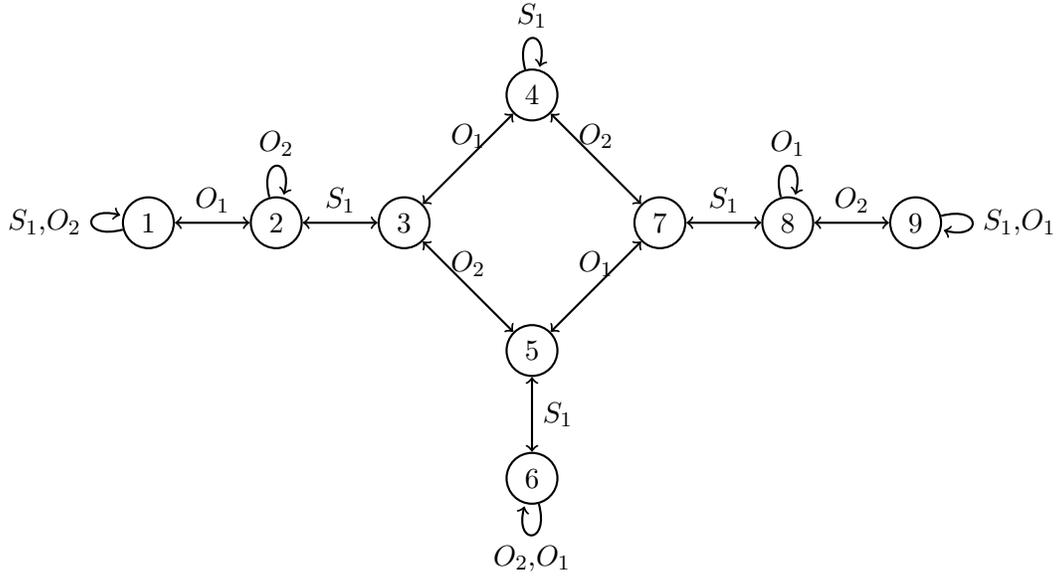
\begin{figure}
\begin{center}
\begin{tikzpicture}[every node,thick] 
 \tikzset{My Style/.style={circle,draw,thick}}

    \node [My Style] (1) at (0,0) {1}; 
    \node [My Style](2) at (1.7,0) {2}; 
    \node [My Style] (3) at (3.4,0) {3};
    \node [My Style](4) at (5.1,1.7) {4};
    \node [My Style](5) at (5.1,-1.7) {5};
    \node [My Style](7) at (6.8,0) {7};
    \node [My Style](8) at (8.5,0) {8};
    \node [My Style](9) at (10.2,0) {9};
    \node [My Style](6) at (5.1,-3.4) {6};


    \draw [<->] (1) --   (2) node [midway,above] {$O_1$}; 
    \draw [<->] (2) -- (3)
    node [midway,above] {$S_1$}; 
    \draw [<->] (3) -- (4)
    node [midway,above] {$O_1$};
    \draw [<->] (3) -- (5)
    node [midway,above] {$O_2$};
    \draw [<->] (4) -- (7)
    node [midway,above] {$O_2$};
    \draw [<->] (5) -- (7)
    node [midway,above] {$O_1$};
    \draw [<->] (5) -- (6)
    node [midway,right] {$S_1$};
    \draw [<->] (7) -- (8)
    node [midway,above] {$S_1$};
    \draw [<->] (8) -- (9)
     node [midway,above] {$O_2$};
    \path[<->] (1)
            edge [loop left] node {$S_1$,$O_2$} ();
    \path[<->](2)
            edge [loop above] node {$O_2$} ();
    \path[<->](8) 
            edge [loop above] node {$O_1$} ();
     \path[<->] (4)
            edge [loop above] node {$S_1$} (); 
    \path[<->] (9)
            edge [loop right] node {$S_1$,$O_1$} ();
    \path[<->] (6)
            edge [loop below] node {$O_2$,$O_1$} ();       

\end{tikzpicture}
\caption{Nine multi-critical CFTs, each one giving rise to a phase diagram with 4 different massive phases as described in the text, are permuted by the action of $O_1$, $O_2$ and $S_1$ as depicted. \label{bosonicCFTs}}
\end{center}
\end{figure}

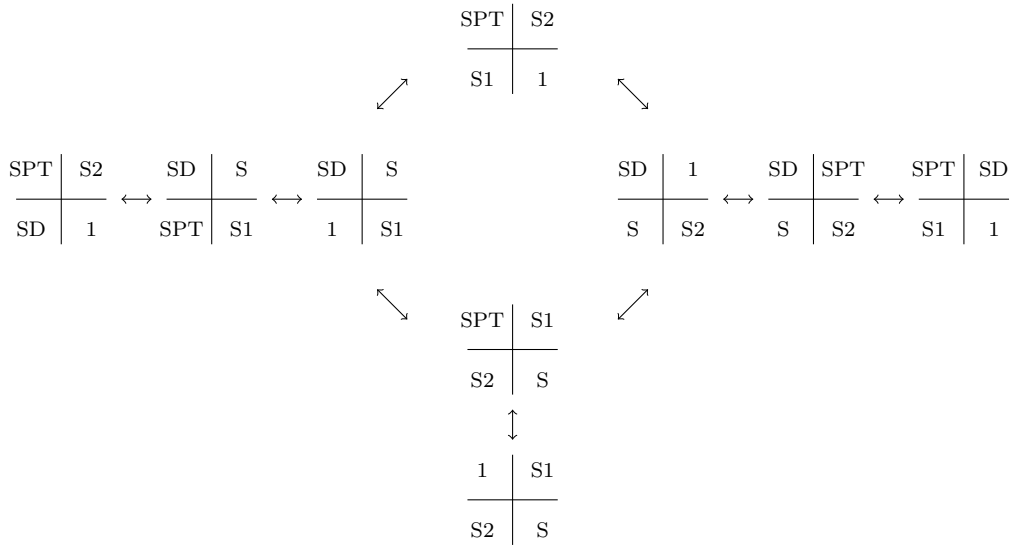
\begin{figure}
\begin{center}

\begin{tikzpicture}
    \begin{scope}
        \draw[-] (-.6,0) -- (.6,0) ;
        \draw[-] (0,-.6) -- (0,.6) ;
        
        \node[font=\scriptsize] at (.4,.4)  {S2};
        \node[font=\scriptsize] at (-.4,.4) {SPT};
        \node[font=\scriptsize] at (-.4,-.4) {SD};
        \node[font=\scriptsize] at (.4,-.4) {1};
    \end{scope}
    
 \begin{scope}[xshift=2cm]
        \draw[-] (-.6,0) -- (.6,0) ;
        \draw[-] (0,-.6) -- (0,.6) ;
        
        \node[font=\scriptsize] at (.4,.4)  {S};
        \node[font=\scriptsize] at (-.4,.4) {SD};
        \node[font=\scriptsize] at (-.4,-.4) {SPT};
        \node[font=\scriptsize] at (.4,-.4) {S1};
    \end{scope}
    
    \begin{scope}[xshift=4cm]
        \draw[-] (-.6,0) -- (.6,0) ;
        \draw[-] (0,-.6) -- (0,.6) ;
        
        \node[font=\scriptsize] at (.4,.4)  {S};
        \node[font=\scriptsize] at (-.4,.4) {SD};
        \node[font=\scriptsize] at (-.4,-.4) {1};
        \node[font=\scriptsize] at (.4,-.4) {S1};
    \end{scope}

     \begin{scope}[xshift=6cm,yshift=2cm]
        \draw[-] (-.6,0) -- (.6,0) ;
        \draw[-] (0,-.6) -- (0,.6) ;
        
        \node[font=\scriptsize] at (.4,.4)  {S2};
        \node[font=\scriptsize] at (-.4,.4) {SPT};
        \node[font=\scriptsize] at (-.4,-.4) {S1};
        \node[font=\scriptsize] at (.4,-.4) {1};
    \end{scope}

     \begin{scope}[xshift=6cm,yshift=-2cm]
        \draw[-] (-.6,0) -- (.6,0) ;
        \draw[-] (0,-.6) -- (0,.6) ;
        
        \node[font=\scriptsize] at (.4,.4)  {S1};
        \node[font=\scriptsize] at (-.4,.4) {SPT};
        \node[font=\scriptsize] at (-.4,-.4) {S2};
        \node[font=\scriptsize] at (.4,-.4) {S};
    \end{scope}

     \begin{scope}[xshift=6cm,yshift=-4cm]
        \draw[-] (-.6,0) -- (.6,0) ;
        \draw[-] (0,-.6) -- (0,.6) ;
        
        \node[font=\scriptsize] at (.4,.4)  {S1};
        \node[font=\scriptsize] at (-.4,.4) {1};
        \node[font=\scriptsize] at (-.4,-.4) {S2};
        \node[font=\scriptsize] at (.4,-.4) {S};
    \end{scope}

     \begin{scope}[xshift=8cm]
        \draw[-] (-.6,0) -- (.6,0) ;
        \draw[-] (0,-.6) -- (0,.6) ;
        
        \node[font=\scriptsize] at (.4,.4)  {1};
        \node[font=\scriptsize] at (-.4,.4) {SD};
        \node[font=\scriptsize] at (-.4,-.4) {S};
        \node[font=\scriptsize] at (.4,-.4) {S2};
    \end{scope}

     \begin{scope}[xshift=10cm]
        \draw[-] (-.6,0) -- (.6,0) ;
        \draw[-] (0,-.6) -- (0,.6) ;
        
        \node[font=\scriptsize] at (.4,.4)  {SPT};
        \node[font=\scriptsize] at (-.4,.4) {SD};
        \node[font=\scriptsize] at (-.4,-.4) {S};
        \node[font=\scriptsize] at (.4,-.4) {S2};
    \end{scope}

     \begin{scope}[xshift=12cm]
        \draw[-] (-.6,0) -- (.6,0) ;
        \draw[-] (0,-.6) -- (0,.6) ;
        
        \node[font=\scriptsize] at (.4,.4)  {SD};
        \node[font=\scriptsize] at (-.4,.4) {SPT};
        \node[font=\scriptsize] at (-.4,-.4) {S1};
        \node[font=\scriptsize] at (.4,-.4) {1};
    \end{scope}
    
   \draw[<->] (.8,0) -- (1.2,0) ;
    \draw[<->] (2.8,0) -- (3.2,0) ;
     \draw[<->] (4.2,1.2) -- (4.6,1.6) ;
 \draw[<->] (7.8,1.2) -- (7.4,1.6);
\draw[<->] (10.8,0) -- (11.2,0) ;
    \draw[<->] (8.8,0) -- (9.2,0) ;
         \draw[<->] (4.2,-1.2) -- (4.6,-1.6) ;
 \draw[<->] (7.8,-1.2) -- (7.4,-1.6);
  \draw[<->] (6.0,-2.8) -- (6.0,-3.2);
    
\end{tikzpicture}
\caption{Phase diagrams for the CFTs from Figure \ref{bosonicCFTs}. To avoid clutter we further abbreviated Tri to 1 and SSB to S. \label{bosonicphasediagrams}}
\end{center}
\end{figure}

The resulting phase diagrams for the original theory of the two Ising scalars, as well as the theories we obtain by acting with $O_1$, $O_2$, and $S_1$, are displayed in figure \ref{bosonicCFTs}. 
CFT$_{SSB}$ and CFT$_{SPT}$ from before correspond to CFTs 6 and 4 respectively.
In total we obtain 9 different CFTs where we only count the CFTs as different if the phase diagrams are qualitatively different; reflections and rotations of the diagram are simply relabelings of the relevant deformations and not genuinely new CFTs. Note in particular that the theory of two Ising scalars we started with is invariant under both the $O_1$ and $O_2$ operation, which simply exchanges pairs of gapped phases. Acting with $S_1$ generates a new CFT. The phase diagrams are:

\begin{eqnarray}
Z_{1} \rightarrow  \left \{ \begin{array}{l} 
Z_{\text{SSB2}}   \cr
Z_{\text{Tri}}\cr 
Z_{\text{SPT}} \cr
Z_{\text{SSBD}}
\end{array} \right . , &\quad& \quad \quad
Z_{2} \rightarrow  \left \{ \begin{array}{l} 
Z_{\text{SSB}} \cr
Z_{\text{SSB1}} \cr 
Z_{\text{SSBD}}  \cr
Z_{\text{SPT}} 
\end{array} \right . , \quad \quad \quad 
Z_{3} \rightarrow \left \{ \begin{array}{l} 
Z_{\text{SSB}}  \cr
Z_{\text{SSB1}}\cr 
Z_{\text{SSBD}} \cr
Z_{\text{Triv}} 
\end{array} \right . , \quad \quad \quad 
\nonumber \\
&& \nonumber \\
&& \nonumber \\
Z_{4} \rightarrow \left \{ \begin{array}{l} 
Z_{\text{SSB2}}  \cr
Z_{\text{Triv}}\cr 
Z_{\text{SPT}} \cr
Z_{\text{SSB1}} 
\end{array} \right . , &\quad& \quad \quad
Z_{5} \rightarrow  \left \{ \begin{array}{l} 
Z_{\text{SSB1}}   \cr
Z_{\text{SSB}}\cr 
Z_{\text{SPT}} \cr
Z_{\text{SSB2}}
\end{array} \right . , \quad \quad \quad
Z_{6} \rightarrow  \left \{ \begin{array}{l} 
Z_{\text{SSB1}}   \cr
Z_{\text{SSB}}\cr 
Z_{\text{1}} \cr
Z_{\text{SSB2}}
\end{array} \right . , \quad \quad \quad
\nonumber \\
&& \nonumber \\
&& \nonumber \\
Z_{7} \rightarrow  \left \{ \begin{array}{l} 
Z_{\text{Triv}} \cr
Z_{\text{SSB2}} \cr 
Z_{\text{SSBD}}  \cr
Z_{\text{SSB}} 
\end{array} \right . , &\quad& \quad \quad
Z_{8} \rightarrow \left \{ \begin{array}{l} 
Z_{\text{SPT}}  \cr
Z_{\text{SSB2}}\cr 
Z_{\text{SSBD}} \cr
Z_{\text{SSB}} 
\end{array} \right . , \quad \quad \quad 
Z_{9} \rightarrow \left \{ \begin{array}{l} 
Z_{\text{SSBD}}  \cr
Z_{\text{Triv}}\cr 
Z_{\text{SPT}} \cr
Z_{\text{SSB1}} 
\end{array} \right . ,
\nonumber \\
&& \nonumber 
\end{eqnarray}
which we summarize in figure \ref{bosonicphasediagrams}.

\vskip10pt

\vskip20pt
\vskip10pt

In Appendix \ref{aA1}, we obtain actions for all nine multi-critical CFTs starting from the action for $\text{Ising}^2$ and applying $O_1$, $O_2$, and $S_1$ as indicated in figure \ref{bosonicCFTs}.  We then deform the mass squareds to verify the phase diagrams in figure \ref{bosonicphasediagrams} explicitly.  

These partial symmetry broken phases where also discussed in \cite{Moradi:2022lqp,Lu:2024ytl}. We can get from the 3 phases related by triality, SPT, Tri and SSB, to the partially broken ones by the action of $O_1$. Conjugating triality with $O_1$ gives rise to a second triality operator, triality'=$O_1$ triality $O_1$, that permutes SSB1, SSB2 and SSBD. The original triality operation permutes CFTs 4, 5, and 6, whereas the action of triality' permutes CFTs 3, 6 and 7. While CFT 6, that is the Ising squared CFT, features in both ``triangles" related by triality, it is a different relevant operator that is turned on in both cases: as is clear from figure \ref{bosonicphasediagrams}, one diagonal of CFT 6 is the Tri to SSB transition featuring in the SPT-SSB-Tri triangle, whereas the other diagonal gives the SSB1 to SSB2 transition required in the second triangle. 

\subsection{Partially Gapped Theories from One Relevant Deformation}
In the previous subsection, we focused on the multi-critical CFTs that govern the phase diagrams we obtained.  These have two relevant parameters, and we verified in Appendix \ref{aA1} that we obtain the correct phases upon appropriately deforming them.  We now examine the situation when we deform one of the two parameters.  This leads to a situation in which one of the two sectors becomes gapped.  There are several situations realized throughout our web of phase diagrams:
\begin{itemize}
    \item The gapped sector is trivial.  This will effectively lead to one copy of the Ising CFT.  These theories lead to transitions between trivial and partially symmetry broken phases and appear in $Z_6$, $Z_7$, $Z_9$, $Z_4$, and $Z_1$.
    \item The gapped sector is an SSB phase.  The resulting theories govern transitions between different SSB phases.  These occur in $Z_6$, $Z_5$, $Z_7$, $Z_8$, $Z_4$, and $Z_2$.
    \item The gapped sector is an SPT phase.  This situation is called a gapless SPT.  It is not intrinsically gapless, since one can deform the mass of the gapless sector to produce an SPT.  Much more about gapless topological phases can be found in i.e. \cite{Verresen:2019igf,Thorngren:2020wet,Wen:2022tkg,Li:2022jbf,Wen:2023otf,Li:2023knf}.  These theories govern transitions between the SPT phase and various SSB phases.  They appear in $Z_5$, $Z_8$, $Z_9$, and $Z_2$.
    \item There are theories in which a dynamical gauge field plays a role that is more subtle than just a Lagrange multiplier or tool to apply Kramers-Wannier duality \footnote{Recall that, following \cite{Karch:2019lnn}, in the terminology of this paper, Kramers-Wannier duality is the statement that Ising is the same theory as $\text{Ising}/\mathbb{Z}_2$.  In terms of actions, this lets us exchange $\int [(D_{a}\phi)^2 + \phi^4 + i\pi aA]$ and $\int [(D_A\phi)^2 + \phi^4]$}.  These can govern all sorts of transitions and appear in $Z_4$, $Z_3$, $Z_2$, and $Z_1$.
\end{itemize}
In Appendix \ref{aA2}, we obtain actions for the partially gapped theories by deforming one of the mass squareds of the CFTs in Appendix \ref{aA1}.  This makes the form of the theories and the phases they govern the transitions between explicit.

\section{Fermionic Theories}
In this section, we discuss the fermionic side of the story told in the previous section.  We begin by discussing the gapped phases with a $\mathbb{Z}_2^F \times \mathbb{Z}_2$ symmetry.  We follow this with a discussion of the transformations between them.  We then describe the web of phase diagrams and the CFTs at their center before turning to the partially gapped theories on the axes of the phase diagrams.  Finally, we describe how to obtain the fermioinic web by fermionizing the bosonic web.

\subsection{Gapped Phases}

Having seen that the duality web techniques can be successfully applied to bosonic phases with a $\mathbb{Z}_2 \times \mathbb{Z}_2$ symmetry, we would like to analyze fermionic theories with a $\mathbb{Z}^F_2 \times \mathbb{Z}_2$ symmetry.  These theories have four topological phases \cite{Tang} which can be nicely described using the SymmTFT framework \cite{Wen:2024udn}. 

A major role in the fermionic theories is played by a $\mathbb{Z}_2$ valued topological invariant, the so called Arf invariant, see \cite{Karch:2019lnn} for an ``all you need to know" introduction to it in the context of the 1+1 duality web. The Arf invariant depends on the boundary conditions of the fermions, and so in particular it depends on the spin structure. As the boundary conditions of charged fermions can be modified by turning on Wilson lines for the corresponding $\mathbb{Z}_2$ gauge fields, we can also included terms which are proportional to Arf$[C \cdot \rho]$, where $\rho$ is the spin structure and $C$ the gauge field (background or dynamical) under which the fermions are charged.

In order to write down the low energy effective theories in terms of the allowed partition functions with $\mathbb{Z}^F_2 \times \mathbb{Z}_2$ symmetry, we need to introduce two background fields $A$ and $C$, where $C$ corresponds to the fermionic symmetry. Only background fields for the fermionic symmetry affect the boundary conditions of fermion fields and so can appear in Arf invariants together with the spin connection. Shifting a fermionic field by a bosonic field gives another fermionic field, so either $C$ or $C+A$ can appear in the Arf.

With this it is easy to write down the partition functions four putative topological phases, the trivial one as well as 3 non-trivial ones. Note that a discussion along these lines appears already in \cite{Huang:2024ror}. We will follow the naming convention chosen there. We believe our systematic field theory treatment elaborates on this previous work in important ways: we construct several CFTs which each sit at the intersection of {\it four} of these phases and in addition show how KT-like transformations exchange these CFTs together with their full phase diagrams, parts of which have been worked out in \cite{Huang:2024ror} as well. 

The partition functions for the non-trivial topological phases are
\begin{eqnarray}
Z_{\text{K}+\text{GW} }[A,C] &=& (-1)^{\text{Arf}[(C+A)\cdot \rho] } \\
 Z_{\text{K}}[A,C] &=&  (-1)^{\text{Arf}[C \cdot \rho]} \\
 Z_{\text{GW}}[A,C] &=& 
 (-1)^{\text{Arf}[(C+A)\cdot \rho]+\text{Arf}[C \cdot \rho]}
 \end{eqnarray}
The subscripts K and GW stand For ``Kitaev" \footnote{So called because it is the low energy description of the nontrivial phase of the Majorana chain in \cite{Kitaev:2000nmw}} and ``Gu-Wen" respectively. One should note that the Arf invariants obey 
\beq
\label{prop}
\text{Arf}[(C+A) \cdot \rho]=
\text{Arf}[C \cdot \rho]+
\text{Arf}[A \cdot \rho]
+\text{Arf}[ \rho] + \int AC
\eeq
so the Arf involving the sum of $A$ and $C$ is ``essentially" the correct way to write a cup product between a fermionic and bosonic background field, mimicking how one obtains topological terms involving $U(1)$ and Spin$_C$ connections in 2+1 dimensions.

In addition, we can spontaneously break the bosonic $\mathbb{Z}_2$ symmetry giving rise to two SSB phases:
\begin{eqnarray}
Z_{\text{SSB}}[A,C] &=&  \delta(A) \\
Z_{\text{SSB+K}}[A,C] &=& \delta(A) (-1)^{\text{Arf}[C \cdot \rho]} 
\end{eqnarray}

\subsection{Transformations Generating the Web}

In \cite{Gaiotto:2020iye}, the authors identify three operations that map $\mathbb{Z}_2^F \times \mathbb{Z}_2$ symmetric spin theories to $\mathbb{Z}_2^F \times \mathbb{Z}_2$ spin theories.  We can gauge the $\mathbb{Z}_2$ symmetry, stack with Arf, or shift the spin structure by a $\mathbb{Z}_2$ gauge field\footnote{They also compose the operations with bosonization. We will return to this later when we connect our fermionic and bosonic CFT webs.}. 
Let's try to put these operations into formulas, very similar to what has been done in \cite{Gaiotto:2020iye}. Following this reference we will refer to the transformations as $S_F$, $O_F$ and $\pi_F$. $S_F$ is stacking with Arf:
\beq
Z[A,C] \xrightarrow[]{\, \,  S_F \, \,} Z[A,C] \, (-1)^{\text{Arf}[\rho \cdot C]}
\eeq
$O_F$ is gauging the bosonic $\mathbb{Z}_2$:
\beq
Z[A,C] \xrightarrow[]{\, \,  O_F \, \,}  \sum_a Z[a,C] \, (-1)^{\int a A}.
\eeq
Last but not least $\pi_F$ shifts the fermionic gauge field by the bosonic one:
\beq
Z[A,C] \xrightarrow[]{\, \,  \pi_F \, \,}  Z[A,C+A] 
\eeq

It is straightforward to work out action of these transformations on the gapped phases. $S_F$ simply adds or removes the corresponding Arf term. In phases without the $\text{Arf}[(A+C) \cdot \rho]$ term, the $O_F$ operation removes a delta function of $A$ if there was one, adds it if there wasn't. To see what $O_F$ does to the remaining two phases we need to calculate:
\begin{eqnarray} \nonumber
\sum_a (-1)^{\text{Arf}[(C+a) \cdot \rho] + \int a A } &=&
\sum_a (-1)^{\text{Arf}[a \cdot \rho] + \int a A + \int CA }
=(-1)^{\text{Arf}[A \cdot \rho] + \text{Arf}[\rho]+ \int CA } \\
&& = (-1)^{\text{Arf}[(C+A) \cdot \rho] + \text{Arf}[C \cdot \rho]}
\end{eqnarray}
Here we first shifted the ``integration variable" by $C$, then used the following identity obeyed by the Arf invariant\footnote{Recall that we absorbed the $2^{-g}$ normalization factor in \cite{Karch:2019lnn} into the definiton of the sum.}, see \cite{Karch:2019lnn}:
\beq \sum_c (-1)^{\text{Arf}[c \cdot{\rho}] + \text{Arf}[\rho]+ \int A c} =(-1)^{\text{Arf}[A \cdot \rho]}   .
\label{id}
\eeq
and last but not least used the property \eqref{prop}. Lo and behold, we see that $O_F$ exchanges GW and GW+K. Last but not least, $\pi_F$ shuffles around the various Arf terms in the SPT phases. In the SSB phases it does not do anything as $A$ is set to zero by the delta functions. All in all the transformations act as summarized in table \ref{fermionicaction}.

\begin{table}
\begin{center}
\begin{tabular}{|c|c|c|c|c|c|c|}
\hline
Original & Tri & K&K+GW&GW&SSB&SSB+K \\
\hline
$S_F$ on Original &K & Tri&GW&K+GW&SSB+K&SSB \\
\hline
$O_F$ on Original &SSB&SSB+K&GW&K+GW&Tri&K \\
\hline
$\pi_F$ on Original & Tri & K+GW & K &  GW & SSB & SSB + K\\
\hline
\end{tabular}
\caption{Action of $O_F$, $S_F$ and $\pi_F$ on the gapped fermionic phases.}
\label{fermionicaction}
\end{center}
\end{table}
\subsection{Transitions and Phase Diagrams}

Probably the simplest CFT one can write down with a $\mathbb{Z}_2^F \times \mathbb{Z}_2$ symmetry is an Ising scalar plus a free Majorana fermion:

\beq
S_{MI} = \int  \left [ (D_A \phi)^2 + \phi^4  +  i \bar{\chi} \slashed{D}_{C \cdot \rho} \chi \right ]
\eeq
This CFT has two relevant deformations, the fermion mass $m$ and the scalar mass squared $M^2$. If both are turned on we realize one of the gapped phases from our list above. The CFT describes the multi-critical point where both relevant operators are turned to zero. The critical lines where only one relevant operator is tuned to zero generically correspond to gapless SPT phases. We'll return to them later. 

It is straightforward to work out the gapped phases this CFT can realize. To do so, we need to recall one more basic fact about the topological terms generated when integrating out a Majorana fermion coupled to the spin connection $C \cdot \rho$: when the fermion has a {\it positive} mass, it simply decouples from the theory an leaves behind a trivial phase, when the fermion has a {\it negative} mass, upon integrating out it generates a non-trivial topological term in the action:
\beq S_{eff} = i \pi \text{Arf}[C \cdot \rho] . \eeq
This is once again akin to what happens with fermions and CS terms in 2+1 dimensions.

For positive mass squared the scalar simply decouples, for negative mass squared it breaks the symmetry. The fermion decouples for either sign of the mass, but for negative mass we generate a non-trivial Arf term. So the phase diagram is as follows:

\beq
\label{mi}
Z_{MI} \xrightarrow[]{m,M^2} \left \{ \begin{array}{lll} 
Z_{\text{Tri}} &\mbox{ for }& m>0, \, M^2>0  \cr
Z_{\text{SSB}}& \mbox{ for }& m>0, \, M^2<0 \cr 
Z_{\text{K}} & \mbox{ for }& m<0, \, M^2>0 \cr
Z_{\text{SSB+K}} & \mbox{ for }& m<0, \, M^2<0
\end{array} \right . , \quad \quad \quad
\eeq

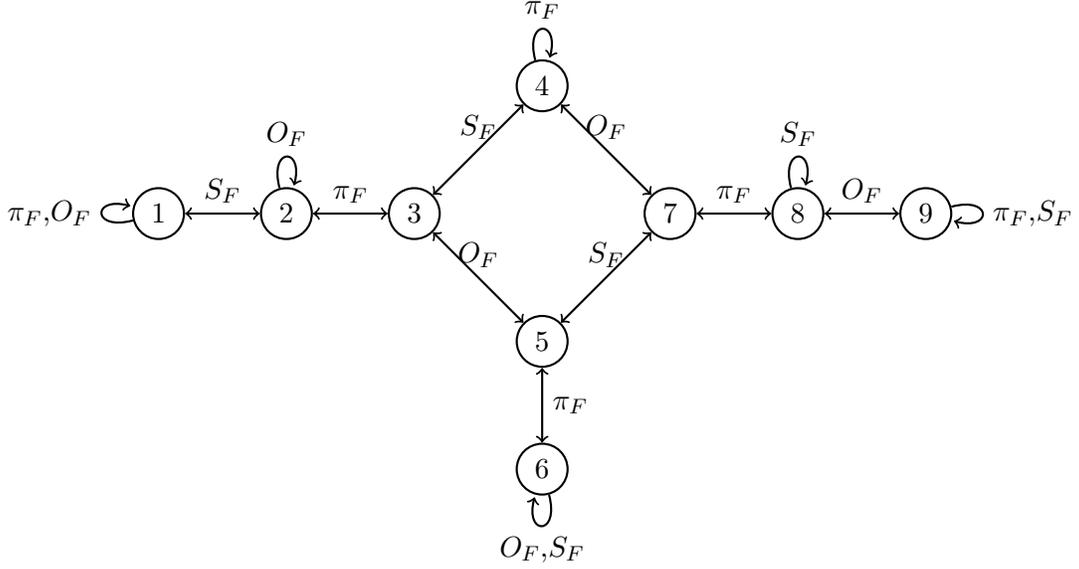
\begin{figure}
\begin{center}
\begin{tikzpicture}[every node,thick] 
 \tikzset{My Style/.style={circle,draw,thick}}

    \node [My Style] (1) at (0,0) {1}; 
    \node [My Style](2) at (1.7,0) {2}; 
    \node [My Style] (3) at (3.4,0) {3};
    \node [My Style](4) at (5.1,1.7) {4};
    \node [My Style](5) at (5.1,-1.7) {5};
    \node [My Style](7) at (6.8,0) {7};
    \node [My Style](8) at (8.5,0) {8};
    \node [My Style](9) at (10.2,0) {9};
    \node [My Style](6) at (5.1,-3.4) {6};


    \draw [<->] (1) --   (2) node [midway,above] {$S_F$}; 
    \draw [<->] (2) -- (3)
    node [midway,above] {$\pi_F$}; 
    \draw [<->] (3) -- (4)
    node [midway,above] {$S_F$};
    \draw [<->] (3) -- (5)
    node [midway,above] {$O_F$};
    \draw [<->] (4) -- (7)
    node [midway,above] {$O_F$};
    \draw [<->] (5) -- (7)
    node [midway,above] {$S_F$};
    \draw [<->] (5) -- (6)
    node [midway,right] {$\pi_F$};
    \draw [<->] (7) -- (8)
    node [midway,above] {$\pi_F$};
    \draw [<->] (8) -- (9)
     node [midway,above] {$O_F$};
    \path[<->] (1)
            edge [loop left] node {$\pi_F$,$O_F$} ();
    \path[<->](2)
            edge [loop above] node {$O_F$} ();
    \path[<->](8) 
            edge [loop above] node {$S_F$} ();
     \path[<->] (4)
            edge [loop above] node {$\pi_F$} (); 
    \path[<->] (9)
            edge [loop right] node {$\pi_F$,$S_F$} ();
    \path[<->] (6)
            edge [loop below] node {$O_F$,$S_F$} ();       

\end{tikzpicture}
\caption{Nine multi-critical CFTs, each one giving rise to a phase diagram with 4 different massive phases as described in the text, are permuted by the action of $S_F$, $O_F$ and $\pi_F$ as depicted. \label{CFTs}}
\end{center}
\end{figure}

Acting on this well understood CFT with $S_F$, $O_F$ and $\pi_F$ we can generate other CFTs, each one mediating a multi-critical transition between 4 of the allowed gapped phases. Once again we only count the CFT as different if the phase diagrams are qualitatively different. This time our starting point, the Majorana + Ising CFT, is invariant under both the S and O operation and only acting with $\pi$ generates a new CFT. The full set of phase diagrams we can realize is depicted in Figure \ref{CFTs}. Once again we get a web of 9 CFTs, where every node depicts an entire phase diagram with CFT and 4 gapped phases. Note that this procedure gives us a full Lagrangian description of the 9 CFTs we obtain this way, as we will spell out and verify in Appendix \ref{aB1}.

\vskip20pt

\vskip10pt

\begin{figure}
\begin{center}

\begin{tikzpicture}
    \begin{scope}
        \draw[-] (-.6,0) -- (.6,0) ;
        \draw[-] (0,-.6) -- (0,.6) ;
        
        \node[font=\scriptsize] at (.4,.4)  {SK};
        \node[font=\scriptsize] at (-.4,.4) {KG};
        \node[font=\scriptsize] at (-.4,-.4) {G};
        \node[font=\scriptsize] at (.4,-.4) {K};
    \end{scope}
    
 \begin{scope}[xshift=2cm]
        \draw[-] (-.6,0) -- (.6,0) ;
        \draw[-] (0,-.6) -- (0,.6) ;
        
        \node[font=\scriptsize] at (.4,.4)  {S};
        \node[font=\scriptsize] at (-.4,.4) {G};
        \node[font=\scriptsize] at (-.4,-.4) {KG};
        \node[font=\scriptsize] at (.4,-.4) {1};
    \end{scope}
    
    \begin{scope}[xshift=4cm]
        \draw[-] (-.6,0) -- (.6,0) ;
        \draw[-] (0,-.6) -- (0,.6) ;
        
        \node[font=\scriptsize] at (.4,.4)  {S};
        \node[font=\scriptsize] at (-.4,.4) {G};
        \node[font=\scriptsize] at (-.4,-.4) {K};
        \node[font=\scriptsize] at (.4,-.4) {1};
    \end{scope}

     \begin{scope}[xshift=6cm,yshift=2cm]
        \draw[-] (-.6,0) -- (.6,0) ;
        \draw[-] (0,-.6) -- (0,.6) ;
        
        \node[font=\scriptsize] at (.4,.4)  {SK};
        \node[font=\scriptsize] at (-.4,.4) {KG};
        \node[font=\scriptsize] at (-.4,-.4) {1};
        \node[font=\scriptsize] at (.4,-.4) {K};
    \end{scope}

     \begin{scope}[xshift=6cm,yshift=-2cm]
        \draw[-] (-.6,0) -- (.6,0) ;
        \draw[-] (0,-.6) -- (0,.6) ;
        
        \node[font=\scriptsize] at (.4,.4)  {1};
        \node[font=\scriptsize] at (-.4,.4) {KG};
        \node[font=\scriptsize] at (-.4,-.4) {SK};
        \node[font=\scriptsize] at (.4,-.4) {S};
    \end{scope}

     \begin{scope}[xshift=6cm,yshift=-4cm]
        \draw[-] (-.6,0) -- (.6,0) ;
        \draw[-] (0,-.6) -- (0,.6) ;
        
        \node[font=\scriptsize] at (.4,.4)  {1};
        \node[font=\scriptsize] at (-.4,.4) {K};
        \node[font=\scriptsize] at (-.4,-.4) {SK};
        \node[font=\scriptsize] at (.4,-.4) {S};
    \end{scope}

     \begin{scope}[xshift=8cm]
        \draw[-] (-.6,0) -- (.6,0) ;
        \draw[-] (0,-.6) -- (0,.6) ;
        
        \node[font=\scriptsize] at (.4,.4)  {K};
        \node[font=\scriptsize] at (-.4,.4) {G};
        \node[font=\scriptsize] at (-.4,-.4) {S};
        \node[font=\scriptsize] at (.4,-.4) {SK};
    \end{scope}

     \begin{scope}[xshift=10cm]
        \draw[-] (-.6,0) -- (.6,0) ;
        \draw[-] (0,-.6) -- (0,.6) ;
        
        \node[font=\scriptsize] at (.4,.4)  {KG};
        \node[font=\scriptsize] at (-.4,.4) {G};
        \node[font=\scriptsize] at (-.4,-.4) {S};
        \node[font=\scriptsize] at (.4,-.4) {SK};
    \end{scope}

     \begin{scope}[xshift=12cm]
        \draw[-] (-.6,0) -- (.6,0) ;
        \draw[-] (0,-.6) -- (0,.6) ;
        
        \node[font=\scriptsize] at (.4,.4)  {G};
        \node[font=\scriptsize] at (-.4,.4) {KG};
        \node[font=\scriptsize] at (-.4,-.4) {1};
        \node[font=\scriptsize] at (.4,-.4) {K};
    \end{scope}
    
   \draw[<->] (.8,0) -- (1.2,0) ;
    \draw[<->] (2.8,0) -- (3.2,0) ;
     \draw[<->] (4.2,1.2) -- (4.6,1.6) ;
 \draw[<->] (7.8,1.2) -- (7.4,1.6);
\draw[<->] (10.8,0) -- (11.2,0) ;
    \draw[<->] (8.8,0) -- (9.2,0) ;
         \draw[<->] (4.2,-1.2) -- (4.6,-1.6) ;
 \draw[<->] (7.8,-1.2) -- (7.4,-1.6);
  \draw[<->] (6.0,-2.8) -- (6.0,-3.2);
    
\end{tikzpicture}
\caption{Phase diagrams for the CFTs from Figure \ref{CFTs}. To avoid clutter we further abbreviated Tri to 1, SSB to S, SSB+K to SK, GW to G and K+GW to KG. \label{phasediagrams}}
\end{center}
\end{figure}
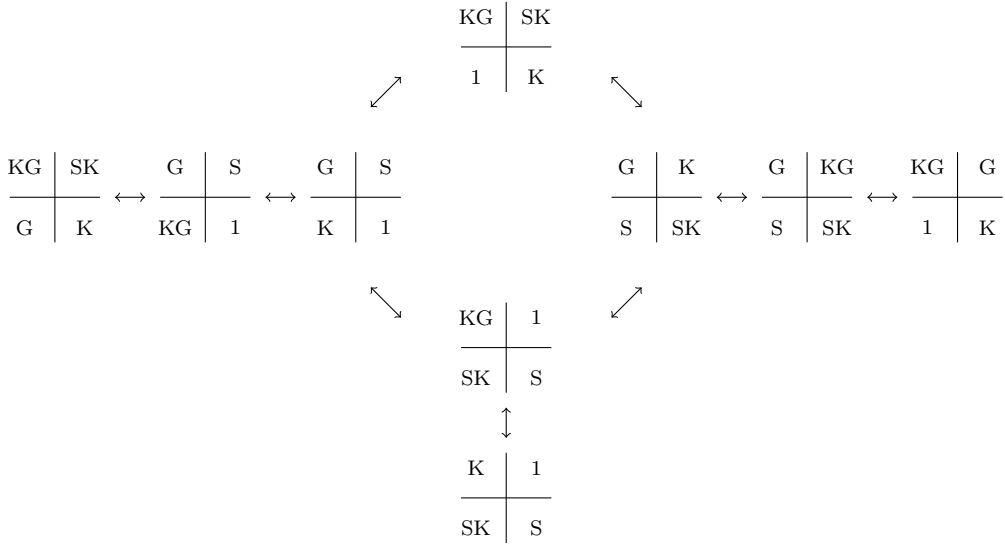

The Ising + Majorana phase diagram of \eqref{mi} is sitting on node 6 in figure \ref{CFTs}. The other CFTs have the following phase diagrams (as in \eqref{mi} these are the gapped phases living in quadrants I, IV, II, III respectively):

\begin{eqnarray}
Z_{1} \rightarrow  \left \{ \begin{array}{l} 
Z_{\text{SSB+K}}   \cr
Z_{\text{K}}\cr 
Z_{\text{K+GW}} \cr
Z_{\text{GW}}
\end{array} \right . , &\quad& \quad \quad
Z_{2} \rightarrow  \left \{ \begin{array}{l} 
Z_{\text{SSB}} \cr
Z_{\text{Tri}} \cr 
Z_{\text{GW}}  \cr
Z_{\text{K+GW}} 
\end{array} \right . , \quad \quad \quad 
Z_{3} \rightarrow \left \{ \begin{array}{l} 
Z_{\text{SSB}}  \cr
Z_{\text{Tri}}\cr 
Z_{\text{GW}} \cr
Z_{\text{K}} 
\end{array} \right . , \quad \quad \quad 
Z_{4} \rightarrow \left \{ \begin{array}{l} 
Z_{\text{SSB+K}}  \cr
Z_{\text{K}}\cr 
Z_{\text{K+GW}} \cr
Z_{\text{Tri}} 
\end{array} \right . ,
\nonumber \\
&& \nonumber \\
&& \nonumber \\
Z_{5} \rightarrow  \left \{ \begin{array}{l} 
Z_{\text{Tri}}   \cr
Z_{\text{SSB}}\cr 
Z_{\text{K+GW}} \cr
Z_{\text{SSB+K}}
\end{array} \right . , &\quad& \quad \quad
Z_{7} \rightarrow  \left \{ \begin{array}{l} 
Z_{\text{K}} \cr
Z_{\text{SSB+K}} \cr 
Z_{\text{GW}}  \cr
Z_{\text{SSB}} 
\end{array} \right . , \quad \quad \quad 
Z_{8} \rightarrow \left \{ \begin{array}{l} 
Z_{\text{K+GW}}  \cr
Z_{\text{SSB+K}}\cr 
Z_{\text{GW}} \cr
Z_{\text{SSB}} 
\end{array} \right . , \quad \quad \quad 
Z_{9} \rightarrow \left \{ \begin{array}{l} 
Z_{\text{GW}}  \cr
Z_{\text{K}}\cr 
Z_{\text{K+GW}} \cr
Z_{\text{Tri}} 
\end{array} \right . ,
\nonumber \\
&& \nonumber 
\end{eqnarray}
We have also summarized these phase diagrams in Figure \ref{phasediagrams}.

Besides the Ising + Majorana CFT another very easy to recognize theory is the one living on node 9: this theory simply describes 2 free Majorana fermions. This is the only CFT that leads to no symmetry breaking phase; if one Majorana couples to $C$ and the other to $C+A$ one simply generates the corresponding Arf invariants whenever the fermion mass is negative.

\vskip20pt
\vskip10pt

A few of those phase transitions have already been identified in \cite{Huang:2024ror}, even though the structure becomes much more apparent when one studies the entire class of phase diagrams we generated. All the transitions in \cite{Huang:2024ror} correspond to transitions going diagonally across the origin in the phase diagram, so that the CFT directly mediates a transition between gapped phases sitting diagonally across from each other. In particular, \cite{Huang:2024ror} already showed that the MI theory can describe a transition between K and SSB (consistent with \eqref{mi}) and the Majorana squared theory between Trivial and GW (as in our $Z_9$). Furthermore \cite{Huang:2024ror} identifies a CFT that can mediate a K to K+GW as well as an SSB+K to trival transition. This is exactly what our $Z_4$ does. Another CFT of \cite{Huang:2024ror} give a SSB+K to GW transition, as does our $Z_8$. Maybe most interestingly, \cite{Huang:2024ror} has a CFT that gives a K+GW to SSB transition. This is something both our $Z_2$ and $Z_8$ can do! To fully specify the CFT clearly looking at just 2 phases is not enough. Since in \cite{Huang:2024ror} this CFT was different from the one of the SSB+K to GW transition, it should not be $Z_8$ again and so should be our $Z_2$.

In Appendix \ref{aB1}, we obtain actions for all nine multi-critical CFTs by starting with $\text{Majorana}+\text{Ising}$ and applying $S_F$, $O_F$, and $\pi_F$ as indicated in \ref{CFTs}.  We then deform the masses and explicitly confirm the phase diagrams in \ref{phasediagrams}.

\subsection{Partially Gapped Theories from Deforming One Relevant Parameter}
Just as we did for the bosonic side of the story, we now examine the effect of deforming one of the relevant parameters in our multicritical CFTs.  As before, the result is a theory with a gapped sector.  Now our possibilities are:
\begin{itemize}
    \item The gapped sector is trivial.  In some cases, these theories are Majorana CFTs.  Depending on the spin structure to which they couple, the either govern the transition between the trivial and Kitaev phases or the trivial and Kiteav+Gu-Wen phases.  These appear in $Z_6$, $Z_5$, $Z_9$, $Z_4$, and $Z_2$.  In other cases, these theories are Ising CFTs that govern the transition between the trivial and SSB phases.  This occurs in $Z_6$, $Z_5$, $Z_3$, and $Z_2$.
    \item The gapped sector spontaneously breaks the bosonic $\mathbb{Z}_2$ symmetry.  The resulting theories govern the transition between the SSB and SSB+Kitaev phases.  They appear in $Z_6$,$Z_5$, $Z_7$, and $Z_8$. 
    \item The gapped sector is in the Kitaev of Kitaev+Gu-Wen phase.  It is tempting to call these gapless SPTs, but the Kitaev phase is not an SPT phase, since it is nontrivial even if we turn of $C$ \footnote{If one desires a more marketable term than ``invertible field theory that is not an SPT because its spin structure dependence ensures it is nontrivial even without a symmetry background," \cite{Ji:2019ugf} refers to the Kitaev phase as an invertible fermionic topological order.}.  There are two situations.  The first is when the gapless sector is Ising.  The resulting theory governs the transition between Kitaev and SSB+Kitaev and appears in $Z_6$, $Z_7$, $Z_4$, and $Z_1$.  The second situation is that the gapless sector is Majorana.  The resulting theory governs the transition between the trivial and Kitaev phases and appears in $Z_9$ and $Z_4$. 
    Gapless topological phases of the form Ising + (Kiteav+Gu-Wen) govern the transition between the SSB + Kitaev and Kitaev + Gu-Wen phases and appears in $Z_5$, $Z_8$, and $Z_4$.  Gapless topological phases of the form Majorana+(Kitaev+Gu-Wen) appear in $Z_8$ and $Z_9$ and govern the transition between Kitaev+Gu-Wen and Gu-Wen.
    \item The gapped sector is in the Gu-Wen phase.  In this case, we have a gapless SPT.  Some examples were found in \cite{Huang:2024ror}.  As before, the gapless SPTs obtained in this way are not intrinsically gapped, since an appropriate mass deformation sends them to an SPT.  Gapless SPT phases of the form Ising + Gu-Wen govern the transition the SSB and Gu-Wen phases and appear in $Z_8$ and  $Z_3$.  
    \item The gapped and gapless sectors can communicate through a dynamical gauge field.  Theories of this sort that govern transitions between fermionic invertible phases (that is Kitaev, Gu-Wen, or Kitaev+Gu-Wen) appear in $Z_3$, $Z_2$, and $Z_1$.  A theory of this sort that controls the phase transition between SSB and Gu-Wen appears in $Z_2$.  A theory of this sort that governs the transition between SSB+Kiteav and Kiteav+Gu-Wen appears in $Z_1$.
\end{itemize}
It's worth noting that there is no transition between the Gu-Wen and trivial phases if one deforms a single parameter!  In Appendix \ref{aB2}, we obtain actions for the partially gapped theories by deforming one of the masses of the CFTs in Appendix \ref{aB1}.  This makes the form of the theories and the phases they govern the transitions between explicit.

\subsection{Relationship to the Bosonic Web}

Both in the bosonic and fermionc case we found structurally equivalent webs with 9 CFTs each. This similarity is not surprising: the two webs can be mapped into each other by bosonization/fermionization.
To flash this out in detail, is useful to briefly discuss the bosonic counterparts of our fermionic gapped phases and the operations we use to map between them.  Recall that, for a bosonic $\mathbb{Z}_2 \times \mathbb{Z}_2$ symmetry, there are six possible gapped phases which we characterize by the partition functions for their ground states:
\beq
    Z_{Tri}[A_1,A_2] = 1,
\eeq
\beq
    Z_{SSB}[A_1,A_2] = \delta(A_1)\delta(A_2),
\eeq
\beq
    Z_{SSB1}[A_1,A_2] = \delta(A_1),
\eeq
\beq
    Z_{SSB2}[A_1,A_2] = \delta(A_2),
\eeq
\beq
    Z_{SSBD}[A_1,A_2] = \delta(A_1+A_2),
\eeq
and
\beq
    Z_{SPT}[A_1,A_2] = (-1)^{\int A_1A_2}.
\eeq
To see what these correspond to in our story, we fermionize with respect to the first $\mathbb{Z}_2$ symmetry.  After renaming a gauge field, this looks like (up to an appropriate normalization)
\beq
    Z_F[C,A] = \sum_{a_1} Z_B [a_1,A] (-1)^{\text{Arf}[a_1 \cdot \rho] + \text{Arf}[\rho] + \int C a_1}. 
\eeq
Thus, we obtain:
\beq
    Z_{F,Tri}[C,A] = \sum_{a_1} (-1)^{\text{Arf}[a_1 \cdot \rho] + \text{Arf}[\rho] + \int C a_1} = (-1)^{\text{Arf}[C \cdot \rho]} = Z_K [C,A],
\eeq
\beq
    Z_{F,SSB} = \sum_{a_1} \delta(a_1)\delta(A) (-1)^{\text{Arf}[a_1 \cdot \rho] + \text{Arf}[\rho\ + \int C a_1} = \delta(A) = Z_{SSB}[C,A],
\eeq
\beq
    Z_{F,SSB1} = \sum_{a_1} \delta(a_1) (-1)^{\text{Arf}[a_1 \cdot \rho] + \text{Arf}[\rho] + \int C a_1} = 1 = Z_{Tri}[C,A],
\eeq
\beq
    Z_{F,SSB2} = \sum_{a_1} \delta(A) (-1)^{\text{Arf}[a_1 \cdot \rho] + \text{Arf}[\rho] + \int C a_1} = \delta(A) (-1)^{\text{Arf}[C \cdot \rho]} = Z_{SSB+K}[C,A],
\eeq
\begin{multline}
    Z_{F,SSBD} = \sum_{a_1} \delta(a_1+A) (-1)^{\text{Arf}[a_1 \cdot \rho] + \text{Arf}[\rho] + \int C a_1} = (-1)^{\text{Arf}[A \cdot \rho] + \text{Arf}[\rho] + \int C A} \\ = (-1)^{\text{Arf}[(C+A)\cdot \rho]+\text{Arf}[C\rho]} = Z_{GW}[C,A],
\end{multline}
and
\beq
    Z_{F,SPT}[C,A] = \sum_{a_1}(-1)^{\text{Arf}[a_1 \cdot \rho]+\text{Arf}[\rho]+\int a_1(A+C)} = (-1)^{\text{Arf}[(A+C) \cdot \rho]} = Z_{K+GW}[C,A].
\eeq
By comparing the composition of a bosonic operation with fermionization, we can see what the fermionic counterparts to bosonic operations are.  Obviously, $O_F$ is the counterpart to gauging the second copy of $\mathbb{Z}_2$.  By noting that gauging the first copy of $\mathbb{Z}_2$ maps us from the trivial bosonic phase to SSB1 and stacking with K maps us from K to Triv, we obtain that $S_F$ is the fermionic counterpart to gauging the first copy of $\mathbb{Z}_2$.  Finally, we note that stacking the SPT maps us from SPT to the trivial bosonic phase and shifting the spin structure maps us from K+GW to K.  Thus, $\pi_F$ is the fermionic counterpart to stacking with SPT \footnote{Or, as some readers might prefer, changing discrete torison.}.  This was also found in \cite{Gaiotto:2020iye}. Table \ref{correspondence} summarizes the correspondence between the bosonic and fermionic gapped phases and operations:

\begin{table}
\begin{center}
\begin{tabular}{|c|c|}
\hline
     Bosonic & Fermionic  \\
\hline
     Tri & K \\
\hline
    SSB & SSB \\
\hline
    SSB1 & Tri \\
\hline
    SSB2 & SSB+K \\
\hline
    SSBD & GW \\
\hline 
    SPT & K+GW \\
\hline
    $O_1$ & $S_F$ \\
\hline
    $O_2$ & $O_F$ \\
\hline
    $S_1$ & $\pi_F$ \\
\hline
\end{tabular}
\end{center}
\caption{Correspondence of the transformations generating the bosonic and fermionic web respectively.}
\label{correspondence}
\end{table}

In \cite{Huang:2024ror}, the author found the same collection of phases and maps between them by fermionizing using the diagonal symmetry.  While this will obviously change the dual symmetry, it leads to the same physics.  All that changes is what fermionizes to what.  

Since, as we just demonstrated, the bosonic and fermionic phases are related by bosonization, we can analyze the action of the triality of \cite{Thorngren:2021yso,Moradi:2022lqp,Lu:2024ytl} on the fermionic phases. The fermionic triality permutes the Kitaev, Kitaev+Gu-Wen, and SSB phases.  The fermionic analog of conjugating the triality operation by $O_1$ is conjugating the fermionic triality by $S_F$.  The resulting triality' permutes the trivial, Gu-Wen, and Kitaev+SSB phases.
\section*{Acknowledgments}
We would like to thank Da-Chuan Lu, Xie Chen and John McGreevy for inspiring talks at the Strings 25 conference as well as the annual UQM Simons collaboration meeting which prompted us to pursue this line of work. We would also like to thank Masaki Oshikawa and David Tong for very helpful comments and discussions. 
This work was supported, in part, by the U.S.~Department of Energy under Grant DE-SC0022021 and by a grant from the Simons Foundation (Grant 651678, AK).

\appendix
\section{Bosonic Actions}
In this appendix, we verify both the phase diagrams and the theories at their center and on their axes by writing and deforming explicit actions for the CFTs in question.  
\subsection{Actions for Multi-Critical CFTs and Verification of the Phase Diagram}
\label{aA1}
We now examine the CFTs in detail to confirm the phase diagrams and the theories on the axes. We start from $Z_6$, which is the $\text{Ising}^2$ CFT, governed by the action
\beq
    S_{6} = \int [(D_{A_1}\phi_1)^2 + (D_{A_2}\phi_1)^2 + \phi_1^4 + \phi_2^4].
\eeq
Since each copy of $\text{Ising}$ governs an SSB transition, we obtain the correct phase diagram.   

Applying $S_1$ maps us to $Z_5$, which has the action:
\beq
    S_5 = \int [(D_{A_1}\phi_1)^2 + (D_{A_2}\phi_1)^2 + \phi_1^4 + \phi_2^4 + i\pi A_1 A_2],
\eeq
which obviously has the correct phase diagram.  

Applying $O_1$ maps us to $Z_7$:
\begin{multline}
    S_7 = \int [(D_{a_1}\phi_1)^2 + (D_{A_2}\phi_2)^2 + \phi_1^4 + \phi_2^4 + i\pi a_1(A_1+A_2)]\\ = \int [(D_{A_1+A_2}\phi_1)^2 + (D_{A_2}\phi_2)^2 + \phi_1^4 + \phi_2^4],
\end{multline}
where the equality is Kramers-Wannier duality.  Taking both masses to positive gives a trivial phase.  Taking the first to be positive and the second to be negative gives SSB2.  Swapping which is positive and which is negative gives SSBD.  Finally, taking both masses to be negative gives SSB.  Thus, we obtain the correct phase diagram. 

Applying $S_1$ maps us to $Z_8$:
\beq
    S_8 = \int [(D_{A_1+A_2}\phi_1)^2 + (D_{A_2}\phi_2)^2 + \phi_1^4 + \phi_2^4 + i\pi A_1 A_2].
\eeq
The phase diagram of this theory simply stacks the SPT on each phase in $S_7$, so it clearly gives the correct phase diagram.  

Applying $O_2$ maps us to $Z_9$:
\beq
    S_9 = \int [(D_{A_1+a_2}\phi_1)^2 + (D_{a_2}\phi_2)^2 + \phi_1^4 + \phi_2^4 + i\pi (A_1 +A_2)a_2].
\eeq
Taking both masses to be positive gives SSBD.  Taking the first to be positive and the second to be negative gives the trivial phase.  Taking the first to be negative and the second to be positive gives SPT.  Taking both to be negative gives SSB1. 

$Z_9$ is a dead end, so return to $Z_7$ and apply $O_2$ to obtain $Z_4$:
\beq
    S_4 = \int[(D_{A_1+a_2}\phi_1)^2 + (D_{a_2}\phi_2)^2+\phi_1^4 + \phi_2^4 +i\pi a_2A_2],
\eeq
Taking both scalar masses to be positive gives SSB2.  Taking the first to be positive and the second to be negative gives the trivial phase.  Taking the first to be negative and the second to be positive gives SPT.  Taking both to be negative gives SSB1.  This is the correct phase diagram. 

We now apply $O_1$ to get to $Z_3$:
\beq
    S_3 = \int[(D_{a_1+a_2}\phi_1)^2 + (D_{a_2}\phi_2)^2+\phi_1^4 + \phi_2^4 +i\pi( a_2A_2 + a_1A_1)]. 
\eeq
Taking the both scalars to have positive masses gives SSB.  Taking the first to be positive and the second to be negative gives SSB1.  Taking the first to be negative and the second to be positive gives SSBD.  Taking both to be negative gives the trivial theory.  This is the correct phase diagram.  

We now apply $S_1$ to get to $Z_2$:
\beq
    S_2 = \int[(D_{a_1+a_2}\phi_1)^2 + (D_{a_2}\phi_2)^2+\phi_1^4 + \phi_2^4 +i\pi (a_2A_2 + a_1A_1 + A_1 A_2)].
\eeq
This clearly just stacks the SPT phase on top of the phase diagram of $S_3$, yielding the correct phase diagram.  

Finally, we can apply $O_1$ to obtain $Z_1$:
\begin{multline}
    S_1 = \int[(D_{a_1+a_2}\phi_1)^2 + (D_{a_2}\phi_2)^2+\phi_1^4 + \phi_2^4 +i\pi (a_2A_2 + a_1s_1 + s_1 A_2 + s_1 A_1)] \\ = \int [(D_{A_1+A_2+a_2}\phi_1)^2 + (D_{a_2}\phi_2)^2 + \phi_1^4 + \phi_2^4 + i\pi a_2A_2].
\end{multline}
Taking both scalar masses to be positive gives SSB2.  Taking the first mass to be negative and the second to be positive gives the SPT phase.  Taking the first mass to be positive and the second to be negative gives the trivial phase.  Taking both to be negative yields SSBD.  
\subsection{Actions for Partially Gapped Theories}
\label{aA2}
We now deform the CFTs in the previous section to these realized explicitly.  We start from $Z_6$, which is the $\text{Ising}^2$ CFT, governed by the action \footnote{We repeat the above actions for convenience.}
\beq
    S_{6} = \int [(D_{A_1}\phi_1)^2 + (D_{A_2}\phi_1)^2 + \phi_1^4 + \phi_2^4].
\eeq
We examine the axes by deforming one parameter at a time.  For positive $M_1^2$, we obtain
\beq
    S_{6}^{M_1^2>0} = \int [(D_{A_2}\phi_1)^2 + \phi_2^4],
\eeq
which is the Ising CFT that governs the transition between the trivial and SSB2 phases.  For negative $M_1^2$, we obtain
\beq
    S_{6}^{M_1^2<0} = \int [(D_{A_2}\phi_1)^2 + \phi_2^4 + i\pi t A_1],
\eeq
which is an $\text{Ising}+\text{SSB1}$ theory that mediates the transition between the SSB and SSB1 phases.  An analogous story happens when we manipulate $M_2^2$.  

$Z_5$ has the action:
\beq
    S_5 = \int [(D_{A_1}\phi_1)^2 + (D_{A_2}\phi_1)^2 + \phi_1^4 + \phi_2^4 + i\pi A_1 A_2].
\eeq
Taking $M_1^2>0$ gives
\beq
    S_5^{M_1^2>0} = \int [(D_{A_2}\phi_1)^2 + \phi_1^4 + i\pi A_1 A_2],
\eeq
which is a known gapless SPT that mediates the transition between the SPT and SSB2 phases.  Taking $M_1^2 < 0$ gives
\beq
    S_5^{M_2^2<0} = \int [(D_{A_2}\phi_1)^2 + \phi_1^4 + i\pi(sA_1+ A_1 A_2)],
\eeq
which is an $\text{Ising}+ \text{SSB2}$ theory that mediates the transition between the SSB1 and SSB phases.  An analogous story unfolds when we deform $M_2^2$.

$Z_7$ has the action:
\beq
    S_7 = \int [(D_{A_1+A_2}\phi_1)^2 + (D_{A_2}\phi_2)^2) + \phi_1^4 + \phi_2^4].
\eeq
If just the first scalar mass is positive, we obtain the Ising CFT that maps between the trivial and SSB2 phases.  If just the second scalar mass is positive, we obtain an Ising CFT that governs the transition between the trivial and SSBD phases.  If we take the first scalar mass squared to be negative, we obtain an $\text{Ising}+ \text{SSBD}$ theory that governs the transition between SSBD and SSB.  If we take the second scalar mass squared to be negative, we obtain an $\text{Ising}+ \text{SSB2}$ that governs the transition between SSB2 and SSB.

$Z_8$ has the action:
\beq
    S_8 = \int [(D_{A_1+A_2}\phi_1)^2 + (D_{A_2}\phi_2)^2 + \phi_1^4 + \phi_2^4 + i\pi A_1 A_2].
\eeq
Taking the first mass squared to be positive gives the gapless SPT,
\beq
    S_8^{M_1^2>0} = \int [(D_{A_2}\phi_2)^2 + \phi_2^4 +i\pi A_1A_2],
\eeq
which governs the transition between the SPT and SSB2 phases.  Taking the second mass squared to be positive gives the gapless SPT,
\beq
    S_8^{M_1^2<0} = \int [(D_{A_1+A_2}\phi_1)^2 + \phi_1^4 +i\pi A_1A_2],
\eeq
which governs the transition between the SPT and SSBD phases.  Taking the first mass squared to be negative gives 
\beq
    S_8^{M_1^2<0} = \int [(D_{A_2}\phi_2)^2 + \phi_2^4 + i\pi( A_1 A_2 + s(A_1+A_2))],
\eeq
which governs the transition between SSBD and SSB.  Taking the second mass squared to be negative gives
\beq
    S_8^{M_2^2<0} = \int [(D_{A_1+A_2}\phi_1)^2 + \phi_1^4 + i\pi( A_1 A_2 + sA_2)],
\eeq
that interpolates between SSB2 and SSB.

$Z_9$ has the action:
\beq
    S_9 = \int [(D_{A_1+a_2}\phi_1)^2 + (D_{a_2}\phi_2)^2 + \phi_1^4 + \phi_2^4 + i\pi (A_1 +A_2)a_2].
\eeq
Taking the first mass squared to be positive yields:
\beq
    S_9^{M_1^2>0} = \int [(D_{a_2}\phi_2)^2 + \phi_2^4 + i\pi(A_1 +A_2)a_2] = \int [(D_{A_1+A_2}\phi_2)^2 + \phi_2^4],
\eeq
which governs the transition between SSBD and the trivial phase.  Taking the second mass squared to be positive gives
\beq
    S_9^{M_2^2>0} = \int [(D_{A_1+a_2}\phi_1)^2 + \phi_1^4 + i\pi (A_1 +A_2)a_2],
\eeq
which governs the transition between the SSBD and SPT phases.  Taking the first mass squared to be negative gives
\begin{multline}
    S_9^{M_1^2<0} = \int [(D_{a_2}\phi_2)^2 + \phi_2^4 + i\pi(A_1 +A_2)a_2 + i\pi s(A_1+a_2)] \\ = \int [(D_{A_1}\phi_1)^2+\phi_1^4 + i\pi A_1A_2],
\end{multline}
which is the gapless SPT that governs the transition between the SPT and SSB1 phases.  Taking the second mass squared to be negative gives
\beq
    S_9^{M_2^2<0} = \int [(D_{A_1+a_2}\phi_1)^2 + \phi_1^4 + i\pi(A_1 +A_2)a_2 + i\pi s(a_2)] = \int [(D_{A_1}\phi_1)^2 + \phi_1^4],
\eeq
which governs the transition between the SSB1 and trivial phases.

$Z_4$ has the action:
\beq
    S_4 = \int[(D_{A_1+a_2}\phi_1)^2 + (D_{a_2}\phi_2)^2+\phi_1^4 + \phi_2^4 +i\pi a_2A_2],
\eeq
Taking the first scalar mass squared to be positive gives
\beq
    S_4^{M_1^2>0} = \int [(D_{a_2}\phi_2)^2 + \phi_2^4 + i\pi a_2A_2] = \int [(D_{A_1}\phi_2)^2 + \phi_2^4],
\eeq
which is an Ising CFT that governs the transition between the trivial and SSB2 phases.  Taking the second mass squared to be positive gives
\beq
    S_4^{M_2^2>0} = \int [(D_{A_1+a_2}\phi_1)^2 + \phi_1^4 + i\pi a_2A_2], 
\eeq
which governs the transition between the SSB2 and SPT phases.  Taking the first mass squared to be negative gives
\begin{multline}
    S_4^{M_1^2<0} = \int [(D_{a_2}\phi_2)^2 + \phi_2^4 + i\pi a_2A_2 + i\pi s(A_1+a_2)] \\ = \int [(D_{A_1}\phi_2)^2 + \phi_2^4 + i\pi A_1 A_2],
\end{multline}
which is the gapless SPT that governs the transition between the SPT and SSB1 phases.  Taking the second mass squared to be negative yields
\beq
    S_4^{M_2^2<0} = \int [(D_{A_1+a_2}\phi_1)^2 + \phi_1^4 + i\pi a_2A_2 + i\pi s a_2]  = \int [(D_{A_1}\phi_1)^2 + \phi_1^4 ],
\eeq
which governs the transition between the trivial and SSB1 phases.

$Z_3$ has the action:
\beq
    S_3 = \int[(D_{a_1+a_2}\phi_1)^2 + (D_{a_2}\phi_2)^2+\phi_1^4 + \phi_2^4 +i\pi( a_2A_2 + a_1A_1)]. 
\eeq
Taking the first scalar mass squared to be positive gives
\beq
    S_3^{M_1^2>0} = \int [(D_{a_2}\phi_2)^2 + \phi_2^4+i\pi( a_2A_2 + a_1A_1)],
\eeq
which governs the transition between the SSB and SSB1 phases.  Taking the second scalar mass squared to be positive gives
\beq
    S_3^{M_2^2>0} = \int [(D_{a_1+a_2}\phi_1)^2 + \phi_1^4 +i\pi( a_2A_2 + a_1A_1)],
\eeq
which governs the transition between the SSB and SSBD phases.  Taking the first scalar mass squared to be negative gives
\begin{multline}
    S_3^{M_1^2<0} = \int [(D_{a_2}\phi_2)^2 + \phi_2^4+i\pi( a_2A_2 + a_1A_1) + i\pi s(a_1+a_2)] \\ = \int [(D_{a_2}\phi_2)^2 + \phi_2^4 + i\pi a_2(A_1+A_2)],
\end{multline}
which controls the phase transition between SSBD and trivial phases.  Taking the second scalar mass squared to be negative gives 
\begin{multline}
    S_3^{M_2^2<0} = \int [(D_{a_1+a_2}\phi_1)^2 + \phi_1^4 +i\pi( a_2A_2 + a_1A_1) + i\pi sa_2] = \int [(D_{a_1}\phi_1)^2 + \phi_1^4 + i\pi a_1 A_1] \\ = \int [(D_{A_1}\phi_1)^2 + \phi_1^4],
\end{multline}
where the first equality comes from integrating out the Lagrange multiplier and the second equality is Kramers-Wannier duality.  This theory governs the transition between the SSB1 and trivial phases.

$Z_2$ has the action:
\beq
    S_2 = \int[(D_{a_1+a_2}\phi_1)^2 + (D_{a_2}\phi_2)^2+\phi_1^4 + \phi_2^4 +i\pi (a_2A_2 + a_1A_1 + A_1 A_2)].
\eeq
Taking the first mass squared to be positive, we obtain
\beq
    S_2^{M_1^2>0} = \int [(D_{a_2}\phi_2)^2 + \phi_2^4 + +i\pi (a_2A_2 + a_1A_1 + A_1 A_2)],
\eeq
which governs the transition between the SSB and SSB1 phases.  Taking the second mass squared to be positive, we obtain
\beq
    S_2^{M_2^2>0} = \int [(D_{a_1+a_2}\phi_1)^2 + \phi_1^4 + +i\pi (a_2A_2 + a_1A_1 + A_1 A_2)],
\eeq,
which controls the transition between the SSB and SSBD phases.  Taking the first mass squared to be negative, we obtain
\begin{multline}
    S_2^{M_1^2<0} = \int [(D_{a_2}\phi_2)^2 + \phi_2^4 + +i\pi (a_2A_2 + a_1A_1 + A_1 A_2+ s(a_1+a_2))]\\ = \int [(D_{a_2}\phi_2)^2 + \phi_2^4 + i\pi a_2(A_1+A_2) + i\pi A_1A_2],
\end{multline}
which governs the transition between SSBD and the SPT phases.  Taking the second mass squared to be negative yields
\begin{multline}
    S_2^{M_2^2<0} = \int [(D_{a_1+a_2}\phi_1)^2 + \phi_1^4 + +i\pi (a_2A_2 + a_1A_1 + A_1 A_2+s a_2)] \\ = \int [(D_{a_1}\phi_1)^2 + \phi_1^4 + i\pi (a_1 A_1 + A_1A_2)] = \int [(D_{A_1}\phi_1)^2 + \phi_1^4 + i\pi A_1A_2],
\end{multline}
which is a gapless SPT that governs the transition between the SSB1 and SPT phases.

$Z_1$ has the action:
\begin{multline}
    S_1 = \int[(D_{a_1+a_2}\phi_1)^2 + (D_{a_2}\phi_2)^2+\phi_1^4 + \phi_2^4 +i\pi (a_2A_2 + a_1s_1 + s_1 A_2 + s_1 A_1)] \\ = \int [(D_{A_1+A_2+a_2}\phi_1)^2 + (D_{a_2}\phi_2)^2 + \phi_1^4 + \phi_2^4 + i\pi a_2A_2].
\end{multline}
Taking the first mass squared to be positive, we obtain
\beq
    S_1^{M_1^2>0} = \int [(D_{a_2}\phi_2)^2+\phi_2^4 +i\pi a_2A_2 ] = \int [(D_{A_2}\phi_2)^2+\phi_2^4 ]
\eeq,
which controls the transition between the trivial and SSB2 phases.  Taking the second mass squared to be positive gives
\beq
    S_1^{M_2^2>0} = \int [(D_{A_1+A_2+a_2} \phi_1)^2 + \phi_1^4 + i\pi a_2A_2], 
\eeq
which governs the transition between the SSB2 and SPT phases.  Taking the first mass squared to be negative gives 
\beq
    S_1^{M_1^2<0}  =\int [(D_{A_1+A_2}\phi)^2 + \phi^4 + i\pi A_1A_2],
\eeq
which controls the transition between the SPT and SSBD phases.  Taking the second mass squared to be negative gives
\beq
    S_1^{M_2^2<0} = \int [(D_{A_1+A_2} \phi)^2 + \phi^4],
\eeq
which is an Ising CFT that governs the transition the trivial and SSBD phases.

\section{Fermionic Actions}
In this appendix, we verify both the phase diagrams and the theories at their center and on their axes by writing and deforming explicit actions for the CFTs in question.  
\subsection{Actions for Multi-Critical CFTs and Verifying the Phase Diagram}
\label{aB1}
We now discuss the phase diagram in detail.  To do so, we start from $\text{Majorana}\boxtimes \text{Ising}$ and apply $S_F$, $O_F$, and $\pi_F$ to obtain explicit Lagrangian descriptions of all nine multicritical CFTs.  We then deform their relevant parameters and verify that we obtain the correct phase diagram.  We begin with $Z_6$, which is $\text{Majorana} + \text{Ising}$, with the action:
\begin{equation}
    S_6 = \int (i\bar{\chi} \slashed{D}_{C \rho} \chi + (D_A \phi)^2 + \phi^4).
\end{equation}
The phase diagram for this CFT is discussed above.

Moving to $Z_5$ is a matter of applying $\pi_F$.  The resulting CFT has the action
\beq
    S_5 = \int (i\bar{\chi} \slashed{D}_{(C+A)\rho} \chi + (D_A \phi)^2 + \phi^4).
\eeq
This indeed has the right phase diagram: the scalar mass takes us from symmetry broken to symmetry unbroken phase, negative fermion mass adds and Arf$[(C+A) \cdot \rho]$, so in total we get Kitaev+Gu-Wen, Trivial, SSB+Kitaev, and SSB as expected. 

Applying $S_F$ moves us to $Z_7$.  The resulting CFT is described by the action
\beq
    S_7 = \int (i\bar{\chi} \slashed{D}_{(C+A)\rho} \chi + (D_A \phi)^2 + \phi^4) + i\pi \text{Arf}[C \cdot \rho]
\eeq
Once again it is easy to see that this has the correct phase diagram. Compared to CFT 5 we add an extra Arf$[C \cdot \rho]$. So the SSB and SSB+Kitaev get exchanged, and Gu-Wen turns into Kitaev+Gu-Wen.

From $Z_7$, we map to $Z_8$ by applying $\pi_F$.  The resulting action is
\beq
    S_8 = \int (i\bar{\chi} \slashed{D}_{C \rho} \chi + (D_A \phi)^2 + \phi^4) + i\pi \text{Arf}[(C+A) \cdot \rho].
\eeq
As expected, negative scalar mass squared gives the two SSB phases with and without the Arf, positive scalar mass squared gives either only the Arf of $C+A$ or both Arf's, that is Kitaev+Gu-Wen and Gu-Wen.  

To map from $Z_8$ to $Z_9$, we apply $O_F$, giving 
\begin{multline}
    S_9 = \int (i\bar{\chi} \slashed{D}_{C\rho} \chi + (D_a \phi)^2 + \phi^4) + i\pi [\text{Arf}[(C+a) \cdot \rho] + \int a A] \\= \int (i\bar{\chi} \slashed{D}_{C\rho} \chi + (D_a \phi)^2 + \phi^4) + i\pi [\text{Arf}[C \cdot \rho ] + \text{Arf}[a \cdot \rho] + \text{Arf}[\rho] + \int a (A+C)]\\ = \int (i\bar{\chi}\slashed{D}_{C\rho} \chi + i\bar{\xi} \slashed{D}_{(A+C)\rho} \xi) + i\pi \text{Arf}[C \cdot \rho] = \int (i\bar{\chi}\slashed{D}_{C\rho} \chi + i\bar{\xi} \slashed{D}_{(A+C)\rho} \xi),
\end{multline}
which describes two free Majorana fermions, as promised.  In the first equality, we use the quadratic property of the Arf invariant.  In the second, we use the fact that fermionizing Ising gives Majorana, and in the third equality, we note that Majorana eats Kitaev \footnote{Somewhat gruesome sounding since both theories are named after people.}.  We now turn to the phase diagram.  If both Majorana masses are positive, we obtain the trivial phase.  If both Majorana masses are negative, we obtain the GW phase.  If one Majorana mass is positive and the other negative, we get either K or GW+K, depending on which mass has which sign.

$Z_9$ is a dead end, so we return to $Z_7$.  Applying $O_F$ maps us to $Z_4$, which has the action 
\beq
    S_4 = \int (i\bar{\chi}\slashed{D}_{(C+a) \cdot \rho}\chi + (D_a \phi)^2 + \phi^4 + i\pi aA ) + i\pi \text{Arf}[C \cdot \rho] 
\eeq
This is our first action in which the dynamical gauge field plays a crucial role.
We can again see the phases easily. For the first time we encounter a phase diagram with only one SSB phase, for which the gauging of the scalar is crucial. Positive fermion and scalar mass squared leaves $a$ as a Lagrange multiplier giving a $\delta(A)$ - the expected SSB+K case. For positive scalar mass squared and negative fermion mass, $a$ appears in an Arf$[(C+a) \cdot \rho]$ term, and so the sum over $a$, using \eqref{id} and \eqref{prop}, gives the sum of the two Arf's and so Kiteav+Gu-Wen once we take the extra Arf$[C \cdot \rho]$ into account. For negative scalar mass squared this time we get a $\delta(a)$ for the dynamical gauge field and so the entire scalar plus flux phase decouples. Depending on fermion mass we get, as expected, Kitaev or the trivial theory.  

To obtain $Z_3$, we apply $S_F$, yielding
\beq
    S_3 = \int (i\bar{\chi}\slashed{D}_{(C+a)\rho}\chi + (D_a \phi)^2 + \phi^4 + i\pi a A).
\eeq
Let us begin by examining the phase diagram.  Positive Majorana and scalar mass lead to an SSB phase.  Positive Majorana mass and negative scalar mass lead to the trivial phase.   Negative Majorana mass and positive scalar mass gives the action $i\pi \text{Arf}[(C+a) \cdot \rho] +i\pi \int aA = i\pi[\text{Arf}[C \cdot \rho] + \text{Arf}[a \cdot \rho] + \text{Arf}[\rho] + \int a(C+A) = i\pi\left [ \text{Arf}[(C+A) \cdot \rho] +\text{Arf}[\rho] \right ]$, so we get the Gu-Wen phase.  Negative Majorana mass and negative scalar mass lead to Kiteav.  

$\pi_F$ maps us to $Z_2$, which has the action
\beq
    S_2 = \int (i\bar{\chi} \slashed{D}_{(C+A+a)\rho}\chi + (D_a \phi)^2 + \phi^4 + i\pi a A) .
\eeq
Let's begin by discussing the phase diagram associated to this CFT.  Positive Majorana and scalar masses deform the theory to the SSB phase.  Negative Majorana mass and positive scalar mass combine to deform the theory to Gu-Wen.  For positive Majorana mass and negative scalar mass, the theory deforms to the trivial theory.  For negative Majorana mass and negative scalar mass, the theory deforms to Kitaev+Gu-Wen.  

Finally, we act with $S_F$ to obtain $Z_1$, whose action is
\beq
    S_1 = \int (i\bar{\chi} \slashed{D}_{(C+A+a)\rho}\chi + (D_a \phi)^2 + \phi^4 + i\pi aA) + i\pi \text{Arf}[C \cdot \rho].
\eeq
Let us briefly verify the phase diagram.  For positive Majorana and scalar mass, the theory deforms to the SSB + K phase.  For positive Majorana mass and negative scalar mass, the theory deforms to the K phase.  For negative Majorana and positive scalar mass, the theory deforms to the GW + K phase.  For negative Majorana and negative scalar mass, the theory deforms to the GW phase.  Thus, it possesses the correct phase diagram.
\subsection{Actions for Partially Gapped Theories}
\label{aB2}
We now show how these possibilities are realized in our collection of CFTs by explicitly deforming one of the parameters and cataloging the results.  We begin with $Z_6$, which is $\text{Majorana} + \text{Ising}$.  It has the action:
\begin{equation}
    S_6 = \int (i\bar{\chi} \slashed{D}_{C \rho} \chi + (D_A \phi)^2 + \phi^4).
\end{equation}
The axis between $\text{Kitaev}$ and $\text{SSB} + \text{Kitaev}$ corresponds to taking the Majorana mass to be negative, in which case we obtain
\begin{equation}
    S^{m<0}_6 = \int ((D_A \phi)^2 + \phi^4) + i\pi \text{Arf}[C \cdot \rho],
\end{equation}
the gapless topological phase $\text{Ising} + \text{Kitaev}$, which mediates the transition between $\text{K}$ and $\text{SSB}+\text{Kitaev}$.  If we instead take the Majorana mass to be positive, we get
\beq
    S^{m>0}_6 = \int ((D_A \phi)^2 + \phi^4)
\eeq
Of course, this is the Ising CFT, which mediates the transition between the $\mathbb{Z}_2$ trivial and $\mathbb{Z}_2$ SSB phases.  Taking $M^2 > 0$, we obtain
\beq
    S^{M^2 > 0}_6 = \int i\bar{\chi} \slashed{D}_{C \rho} \chi,
\eeq
which is simply the Majorana CFT that mediates the transition between the trivial and Kitaev phases.  Taking $M^2 < 0$ gives
\beq
    S^{M^2 < 0}_6 = \int  (i\bar{\chi} \slashed{D}_{C \rho} \chi + i\pi sA),
\eeq
where $s$ is an auxillary $\mathbb{Z}_2$ gauge field that can be integrate out to give a delta function in the path integral.  Thus, this theory is the product of the Majorana CFT and a $\mathbb{Z}_2$ SSB phase.  It mediates the transition between SSB and $\text{SSB} + \text{Kitaev}$.

$Z_5$ has the action
\beq
    S_5 = \int (i\bar{\chi} \slashed{D}_{(C+A)\rho} \chi + (D_A \phi)^2 + \phi^4).
\eeq 
The axis between $\text{Kitaev} + \text{Gu-Wen}$ and $\text{SSB} + \text{Kitaev}$ again corresponds to negative Majorana mass, so the relevant theory is 
\beq
    S^{m<0}_5 = \int ((D_A \phi)^2 + \phi^4) + i\pi \text{Arf}[(C+A) \cdot \rho],
\eeq
which describes the gapless topological phase $\text{Ising} + (\text{Kitaev} + \text{Gu-Wen})$.  Note that since $\pi_F$ only affects the spin portion of the theory, we could have simply applied it to $\text{Ising} + \text{Kitaev}$.  If we instead consider positive Majorana mass, we again obtain the Ising CFT, which mediates the transition between the SSB and trivial phases.  If we consider $M^2 > 0$, we obtain
\beq
    S_5^{M^2 > 0} = \int (i\bar{\chi}\slashed{D}_{(C+A)\rho}\chi),
\eeq
which is a Majorana CFT that mediates the transition between the trivial and Kitaev + Gu-Wen phases.  Note that this CFT is coupled to a different spin structure than the one described by $S^{M^2>0}_6$.  Finally, 
\beq
    S_5^{M^2 < 0} = \int (i\bar{\chi}\slashed{D}_{(C+A)\rho}\chi + i\pi s A), 
\eeq
which is a $\text{SSB}+ \text{Majorana}$ CFT that mediates the transition between the SSB and SSB + Kitaev phases.

$Z_7$ has by the action
\beq
    S_7 = \int (i\bar{\chi} \slashed{D}_{(C+A)\rho} \chi + (D_A \phi)^2 + \phi^4) + i\pi \text{Arf}[C \cdot \rho]
\eeq
We can turn on a positive Majorana mass, yielding the action
\beq
     S_7^{m>0} = \int((D_A \phi)^2 + \phi^4)) + i\pi \text{Arf}[C \cdot \rho] ,
\eeq
which describes the gapless topological $\text{Ising}+\text{Kitaev}$ mediating the transition between Kitaev and SSB+Kitaev.  Turning on a negative Majorana mass gives
\beq
    S_7^{m<0} = \int((D_A \phi)^2 + \phi^4)) + i\pi \text{Arf}[(C+A) \cdot \rho] + i\pi \text{Arf}[C \cdot \rho],
\eeq
which describes the gapless SPT the mediates the phase transition between the SSB and Gu-Wen phases.  We can take $M^2 > 0$ to obtain
\beq
    S_7^{M^2 > 0} = \int (i\bar{\chi}\slashed{D}_{(C+A)\rho}\chi) + i\pi \text{Arf}[C \cdot \rho],
\eeq
which mediates the phase transition between Kiteav and Gu-Wen.  Finally, turning on a positive scalar mass yields
\beq
    S_7^{M^2 < 0} = \int (i\bar{\chi}\slashed{D}_{(C+A)\rho}\chi + i\pi sA) + i\pi \text{Arf}[C \cdot \rho],
\eeq
which mediates the phase transition between the SSB + Kitaev and SSB phases.

$Z_8$ has the action
\beq
    S_8 = \int (i\bar{\chi} \slashed{D}_{C \rho} \chi + (D_A \phi)^2 + \phi^4) + i\pi \text{Arf}[(C+A) \cdot \rho].
\eeq
We begin our investigation of the theories on the axes by turning on a positive Majorana mass, yielding
\beq
    S^{m>0}_8 = \int ((D_A \phi)^2 + \phi^4) +i\pi \text{Arf}[(C+A) \cdot \rho],
\eeq
which describes the gapless topological phase $\text{Ising}+(\text{K}+\text{GW})$ that mediates the phase transition between SSB + Kitaev and Kitaev + Gu-Wen.  If we instead turn on a negative Majorana mass, we obtain
\beq
    S^{m<0}_8 = \int ((D_A \phi)^2 + \phi^4) +i\pi \text{Arf}[(C+A) \cdot \rho] + i\pi \text{Arf}[C \cdot \rho],
\eeq
which is the $\text{Ising} + \text{GW}$ gapless SPT that mediates the transition between the SSB and Gu-Wen phases.  If we turn on a positive $M^2$, we obtain
\beq
    S^{M^2>0}_8 = \int i\bar{\chi} \slashed{D}_{C \rho} \chi + i\pi \text{Arf}[(C+A) \cdot \rho],
\eeq
which is a gapless topological phase $\text{Majorana}+(\text{GW}+\text{K})$ that controls the phase transition between Gu-Wen and Gu-Wen+Kitaev.  Finally, a negative $M^2$ begets
\beq
    S^{M^2<0}_8 = \int (i\bar{\chi} \slashed{D}_{C \rho} \chi + i\pi sA) + i\pi \text{Arf}[(C+A) \cdot \rho],
\eeq
which mediates the transition between the SSB+Kitaev and SSB phases.

$Z_9$ has the action
\beq
    S_9 = \int (i\bar{\chi}\slashed{D}_{C\rho} \chi + i\bar{\xi} \slashed{D}_{(A+C)\rho} \xi),
\eeq
We can take the first Majorana mass to be negative, yielding
\beq
    S^{m_1<0}_9 = \int (i\bar{\chi} \slashed{D}_{C\rho} \chi) + i\pi \text{Arf}[(A+C) \cdot \rho],
\eeq
describing the gapless topological phase $\text{Majorana}+ (\text{K}+\text{GW})$, which mediates the transition between Gu-Wen and Kitaev + Gu-Wen.  Flipping the sign of the mass gives
\beq
    S^{m_1>0}_9 = \int (i\bar{\chi} \slashed{D}_{C\rho} \chi),
\eeq
which is a Majorana CFT describing the transition between the trivial and Kitaev phases.  We can instead turn on the other Majorana mass.  A negative $m_2$ provides
\beq
    S^{m_2<0}_9 = \int (i\bar{\xi} \slashed{D}_{(C+A)\rho} \xi) + i\pi \text{Arf}[C \cdot \rho],
\eeq
which describes the gapless topological phase $\text{Majorana}+ \text{Kitaev}$ that mediates the phase transition between Gu-Wen and Kitaev.  The final axis contains the theory described by the action
\beq
    S_9^{m_2>0} = \int i\bar{\xi} \slashed{D}_{(C+A)\rho} \xi,
\eeq
which is again the Majorana CFT that mediates the transition between Gu-Wen + Kitaev and the trivial phase.

$Z_4$ has the action 
\beq
    S_4 = \int (i\bar{\chi}\slashed{D}_{(C+a) \cdot \rho}\chi + (D_a \phi)^2 + \phi^4 + i\pi aA ) + i\pi \text{Arf}[C \cdot \rho] 
\eeq
Let's explore what's on the axes by turning on one deformation at a time.  If we turn on a positive Majorana mass, we have
\beq
    S^{m>0}_4 = \int ((D_a \phi)^2 + \phi^4 + i\pi aA) + i\pi \text{Arf}[C \cdot \rho] = \int ((D_A\phi)^2 + \phi^4) + i\pi \text{Arf}[C \cdot \rho],
\eeq
where the second equality is Kramers-Wannier duality.  This is the gapless topological phase $\text{Ising}+ \text{Kitaev}$ that mediates the transition between SSB + Kitaev and Kitaev.  Flipping the sign of the mass provides 
\begin{multline}
    S^{m<0}_4 = \int ((D_a \phi)^2 + \phi^4 + i\pi aA ) + i\pi \text{Arf}[C \cdot \rho] + i\pi \text{Arf}[(C+a) \cdot \rho] \\ = \int ((D_a \phi)^2 + \phi^4 + i\pi a(A+C)) + i\pi\text{Arf}[a \cdot \rho] + i\pi \text{Arf}[\rho] \\ = \int i\bar{\chi}\slashed{D}_{(A+C)\rho}\chi,
\end{multline}
which is a Majorana CFT that describes the transition between the trivial and K+GW phases. 
 Turning on a positive scalar mass gives
\begin{multline}
    S_4^{M^2>0} = \int (i\bar{\chi}\slashed{D}_{(C+a) \cdot \rho}\chi + i\pi aA) + i\pi \text{Arf}[C \cdot \rho] \\ = \int (i\bar{\chi}\slashed{D}_{a' \cdot \rho}\chi + i\pi (a' + C)A) + i\pi \text{Arf}[C \cdot \rho] \\ = \int (i\bar{\chi}\slashed{D}_{a' \cdot \rho}\chi + i\pi aA) + i\pi \text{Arf}[(C+A) \cdot \rho] + i\pi \text{Arf}[A \cdot \rho] + i\pi \text{Arf}[\rho] \\ = \int ((D_A\phi)^2 + \phi^4) + i\pi \text{Arf}[(C+A) \cdot \rho].
\end{multline}
Above, the second equality comes from relabeling integration variables, the third equality comes from the quadratic property of the Arf invariant, and the fourth equality comes from the fact that bosonizing the Majorana fermion gives the Ising model.  The resulting theory is the gapless topological phase $\text{Ising}+(\text{Kitaev}+\text{Gu-Wen})$ that mediates the phase transition between SSB + Kitaev and Kitaev + Gu-Wen.  Finally, turning on a negative scalar mass gives
\beq
    S_4^{M^2<0} = \int i\bar{\chi}\slashed{D}_{C\rho}\chi + i\pi\text{Arf}[C \cdot \rho],
\eeq
which is the gapless topological phase $\text{Majorana}+ \text{Kitaev}$.  It controls the transition between the trivial and Kitaev phases.

$Z_3$ has the action
\beq
    S_3 = \int (i\bar{\chi}\slashed{D}_{(C+a)\rho}\chi + (D_a \phi)^2 + \phi^4 + i\pi a A).
\eeq
We move to the theories on the axes.  We begin by considering massive fermions.  With positive mass, we obtain
\beq
    S_3^{m>0} = \int ((D_a \phi)^2 + \phi^4 + i\pi aA) = \int ((D_A \phi)^2 + \phi^4),
\eeq
which is the Ising CFT that mediates the phase transition between SSB and trivial phases.  A negative Majorana mass gives
\begin{multline}
    S_3^{m<0} = \int ((D_a \phi)^2 + \phi^4 + i\pi aA) + i\pi \text{Arf}[(C+a) \cdot \rho]\\
    =\int ((D_a \phi)^2 + \phi^4 + i\pi aA) + i\pi[\text{Arf}[C \cdot \rho] +\text{Arf}[a \cdot \rho]+\text{Arf}[\rho] + \int aC],
\end{multline}
which governs the transition between Gu-Wen and Kitaev phases.  Now we endow the scalars with mass.  For positive mass, we obtain
\begin{multline}
    S_3^{M^2 >0} = \int (i\bar{\chi}\slashed{D}_{(C+a)\rho}\chi + i\pi a A) = \int (i\bar{\chi}\slashed{D}_{a'\rho}\chi + i\pi (a'+C) A) \\ =\int ((D_A\phi)^2 + \phi^4 +i\pi CA) + i\pi[\text{Arf}\left [ A \cdot \rho] + \text{Arf}[\rho] \right ] \\ =\int ((D_A\phi)^2 + \phi^4) + i\pi\left [ \text{Arf}[(C+A) \cdot \rho] +\text{Arf}[C \cdot \rho] \right ].
\end{multline}
Above, the second equality follows from relabeling an integration variable, the second from bosonizing the Majorana theory, and the third from the quadratic property of the Arf invariant.  The result is the gapless SPT $\text{Ising}+ \text{Gu-Wen}$ that mediates the transition between the Gu-Wen and SSB phases.  For negative mass, we obtain
\beq
    S_3^{M^2<0} = \int (i\bar{\chi} \slashed{D}_{(C+a)\rho}\chi + i\pi Aa + sa)) = \int [i\bar{\chi} \slashed{D}_{(C)\rho}\chi],
\eeq
which is a Majorana CFT that governs the transition between the trivial and Kitaev phases.

$Z_2$ has the action
\beq
    S_2 = \int (i\bar{\chi} \slashed{D}_{(C+A+a)\rho}\chi + (D_a \phi)^2 + \phi^4 + i\pi a A) .
\eeq
Let's now turn on one relevant deformation at a time.  For positive Majorana mass, we obtain
\beq
    S_2^{m>0} = \int ((D_a\phi)^2 + \phi^4 + i\pi aA) = \int ((D_A \phi)^2 + \phi^4),
\eeq
which is the Ising CFT that governs the transition between the SSB and trivial phase.  For negative Majorana mass, we obtain:
\beq
    S_2^{m<0} = \int ((D_a\phi)^2 + \phi^4 + i\pi aA) + i\pi \text{Arf}[(A+C+a) \cdot \rho],
\eeq
which controls the phase transition between Kitaev + Gu-Wen and Gu-Wen.  For positive scalar mass, we obtain
\beq
    S_2^{M^2>0} = \int (i\bar{\chi} \slashed{D}_{(C+A+a)\rho}\chi + i\pi a A),
\eeq
which controls the transition between SSB and Gu-Wen phases.  For negative scalar mass,
\beq
    S_2^{M^2<0} = \int i \bar{\chi} \slashed{D}_{(C+A)\rho}\chi,
\eeq
which governs the transition between Gu-Wen + Kitaev and trivial phases.

$Z_1$ has the action:
\beq
    S_1 = \int (i\bar{\chi} \slashed{D}_{(C+A+a)\rho}\chi + (D_a \phi)^2 + \phi^4 + i\pi aA) + i\pi \text{Arf}[C \cdot \rho].
\eeq
We now move to deforming parameters.  For positive Majorana mass, we have
\beq
    S_1^{m>0} = \int((D_a\phi)^2 + \phi^4 + i\pi aA) + i\pi\text{Arf}[C \cdot \rho] = \int ((D_A\phi)^2 + \phi^4) + i\pi\text{Arf}[C \cdot \rho].
\eeq
This is the gapless topological phase $\text{Ising}+ K$ that presides over the phase transition between Kitaev and SSB+Kitaev.  For negative Majorana mass, we obtain 
\beq
    S_1^{m<0} = \int ((D_a\phi)^2 + \phi^4 + i\pi aA) + i\pi \text{Arf}[(C+A+a) \cdot \rho] + i\pi\text{Arf}[C \cdot \rho],
\eeq
which governs the transition between Gu-Wen and Gu-Wen+Kitaev (to show the latter, one needs to invoke the quadratic property of the Arf invariant and integrate out $a$).  For positive scalar mass, we obtain
\beq
    S_1^{M^2>0} = \int (i\bar{\chi}\slashed{D}_{(C+A+a)\rho}\chi + i\pi aA) + i\pi \text{Arf}[C \cdot \rho],
\eeq
which controls the transition between SSB+Kitaev and Kitaev + Gu-Wen (to show the latter, expand the Arf invariant using the quadratic property and integrate out $a$).  Finally, negative scalar mass gives
\beq
    S_1^{M^2<0} = \int i\bar{\chi}\slashed{D}_{(C+A)\rho}\chi + i\pi \text{Arf}[C \cdot \rho],
\eeq
which describes the transition between Kitaev and Gu-Wen.

\bibliographystyle{JHEP}
\bibliography{kennedytasaki}

\end{document}